\documentclass[authoryear,12pt]{article}

\usepackage{natbib}
\usepackage[title,titletoc]{appendix}
\usepackage[english]{babel}
\usepackage{amsmath}
\usepackage{latexsym}
\usepackage{amssymb}
\usepackage{amsmath}
\usepackage{amsthm}
\usepackage{mathtools}
\usepackage{multirow}
\usepackage{multicol}
\usepackage{float}
\usepackage{graphicx}
\usepackage{booktabs}
\usepackage{hyperref}
\usepackage{array}
\usepackage{hyperref}
\usepackage{enumitem}
\usepackage{colortbl}
\usepackage{subfig}
\usepackage{epsfig}
\usepackage{lscape}
\usepackage[normalem]{ulem}
\usepackage[usenames,dvipsnames]{xcolor}
\usepackage{tikz}
\usetikzlibrary{bayesnet}
\usepackage{bbm}
\usepackage{caption} 
\usepackage{rotating}
\usepackage{algorithm}
\usepackage{algpseudocode}
\usepackage[blocks]{authblk} 
\makeatletter
\newcommand*{\centernot}{%
	\mathpalette\@centernot
}
\def\@centernot#1#2{%
	\mathrel{%
		\rlap{%
			\settowidth\dimen@{$\m@th#1{#2}$}%
			\kern.5\dimen@
			\settowidth\dimen@{$\m@th#1=$}%
			\kern-.5\dimen@
			$\m@th#1\not$%
		}%
		{#2}%
	}%
}
\makeatother

\newcommand{\vertiii}[1]{{\left\vert\kern-0.20ex\left\vert\kern-0.20ex\left\vert #1 
	\right\vert\kern-0.20ex\right\vert\kern-0.20ex\right\vert}}
	
	\newcommand{\norm}[1]{\left\lvert#1\right\rvert}
	\newcommand{\lnorm}[1]{\left\lVert#1\right\rVert}

	\newtheorem{defin}{Definition}

	\newtheorem{theorem}{Theorem}
	\newtheorem{corollary}{Corollary}
	\newtheorem{proposition}{Proposition}
	
	\newtheorem{lemma}{Lemma}

\newcounter{example}
\newtheorem{ex}[example]{Example}

\newcommand{\N}{\mathbb{N}}
\newcommand{\Z}{\mathbb{Z}}
\newcommand{\R}{\mathbb{R}}
\newcommand{\E}{\mathrm{E}}

\newcommand{\V}{\mathrm{V}}

\newcommand{\Fb}{\mathcal{F}}
\newcommand{\Ob}{\mathcal{O}}

\hypersetup{
colorlinks=true,
linkcolor=blue,
filecolor=blue,      
urlcolor= blue,
citecolor=blue,
}

\DeclareMathOperator*{\argmax}{arg\,max}

\begin{document}


\title{Pseudo-variance quasi-maximum likelihood estimation of  semi-parametric  time series models\footnote{Corresponding author: Paolo Gorgi. Email address: p.gorgi@vu.nl.}}

\author[1]{Mirko Armillotta}
\author[2,3]{Paolo Gorgi}

\affil[1]{University of Rome Tor Vergata, Rome, Italy}
\affil[2]{Vrije Universiteit Amsterdam, Amsterdam, The Netherlands} \affil[3]{Tinbergen Institute, Amsterdam, The Netherlands}

\date{November 13, 2024}

{
	\makeatletter
	\renewcommand\AB@affilsepx{: \protect\Affilfont}
	\makeatother
	
	
	\makeatletter
	\renewcommand\AB@affilsepx{, \protect\Affilfont}
	\makeatother
	
}

{\let\newpage\relax\maketitle}

\vspace{-0.1cm}
\begin{abstract}
	{\small		
		We propose a novel estimation approach for a general class of semi-parametric time series models where the conditional expectation is modeled through a parametric function. The proposed class of estimators  is based on a Gaussian quasi-likelihood function and it relies on the specification of a parametric pseudo-variance  that can contain parametric restrictions with respect to the conditional expectation. The specification of the pseudo-variance and the parametric  restrictions  follow naturally in observation-driven models with  bounds in the support of the observable process, such as count processes and double-bounded time series. We derive the asymptotic properties of the estimators and   a validity test for the  parameter restrictions. We show that the results remain valid irrespective of the correct specification of the pseudo-variance. The key advantage of the restricted estimators is that they can achieve higher efficiency compared to alternative quasi-likelihood methods that are available  in the literature. Furthermore, the testing approach can be used to build specification tests for parametric time series models.  We illustrate the practical use of the methodology in a simulation study and two empirical applications featuring integer-valued autoregressive processes, where  assumptions on the dispersion of the thinning operator are formally tested,  and autoregressions for double-bounded data  with application to a realized correlation time series.
		
	}		
\end{abstract}

\noindent	\emph{Keywords:} Double-bounded time series, integer-valued autoregressions, quasi-maximum likelihood.\\
\noindent	\emph{JEL codes:} C32, C52, C58.\\


\section{Introduction} 
\label{SEC: Introduction}

A wide range of time series models have been proposed  in the literature to model the conditional mean of time series data. Their specification often depends on   the nature of the time series  variable of interest.  For example, AutoRegressive Moving Average (ARMA) models \cite[]{box1970time} are typically employed for  time series variables that are continuous  and take values on the real line.     INteger-valued AutoRegressive (INAR) models  \cite[]{al1987, MCK1988} and   INteger-valued GARCH  models (INGARCH) \cite[]{Heinen(2003), fer2006}  are designed to account for the discrete and non-negative nature  of count processes. 
Autoregressive Conditional Duration (ACD) models  \cite[]{engle1998autoregressive} are used   for modeling   non-negative  continuous  processes.    Beta autoregressive models \cite[]{rocha2009beta}  are employed for modeling  double-bounded time series data lying in a specified interval domain.    The estimation of such models can be carried out by  the Maximum Likelihood Estimator (MLE), which  constitutes the gold standard approach for the estimation of unknown parameters in parametric models.  However, the MLE requires parametric assumptions on the entire conditional distribution of the time series process. This feature is not appealing when the interest of the study  is only on modeling the conditional mean instead of the entire conditional distribution. Furthermore, the likelihood function can sometimes present a complex form  and the implementation of the MLE can become  unfeasible. For instance, maximum likelihood inference of INAR($p$) models  is well-known to be cumbersome and numerically difficult  when the order of the model $p$ is large \citep{pedeli2015likelihood}. In such situations, the use of quasi-likelihood methods becomes attractive.

The   Quasi-MLE (QMLE), introduced by \cite{wedderburn1974quasi}, is a likelihood-based estimator where there is a quasi-likelihood  that is not necessarily the true distribution of the data. Quasi-likelihoods are typically a member of the one-parameter exponential family. \cite{gourieroux1984pseudo} show that the QMLE is consistent for the true unknown parameters of the model. Nevertheless, QMLEs can be inefficient because, given a parametric definition for the conditional mean of the process, the  conditional variance is implicitly  constrained to be a function of the conditional mean as determined by the  exponential family of distributions that is considered. In order to improve the estimation efficiency  for the parameters of the conditional mean in time series models,  \cite{francq_2021two} propose a two-stage Weighted Least Squares Estimator (WLSE) where in the first step the conditional variance of the process is estimated and it is then used in the second step as weighting sequence for the solution of the weighted least squares problem. It is shown that this WLSE leads to improved efficiency with respect to QMLE if the variance function is correctly specified. A similar estimator has been more recently proposed in the context of estimating functions approach leading to the same type of efficiency improvement \citep{francq2023optimal}. 

In this paper, we  propose  a novel class  of  QMLEs for the estimation of the conditional expectation of semi-parametric time series models. The estimators are based on a Gaussian quasi-likelihood and a pseudo-variance specification, which can contain restrictions with the parameters of the conditional expectation.  The Pseudo-Variance  QMLEs (PVQMLEs) only require   parametric assumptions on the conditional expectation as the  pseudo-variance function  does not need  to be correctly specified.  We establish strong consistency and asymptotic normality of the PVQMLEs under very general conditions. The case in which the pseudo-variance formulation corresponds to the true conditional variance of the process is obtained as a special case. We show that when no restrictions are imposed between the mean and pseudo-variance, the resulting unrestricted PVQMLE has the same asymptotic efficiency of a particular WLSE. Furthermore, if the pseudo-variance is correctly specified it achieves the same asymptotic efficiency as the efficient WLSE.  On the other hand, when parameter restrictions are  considered, the resulting restricted PVQMLEs can achieve higher efficiency compared to the efficient WLSE and alternative QMLEs.  This result is theoretically shown in some special cases and   empirically verified for INAR models through an extensive numerical exercise. We discuss how the specification of the pseudo-variance and the parameter restrictions naturally arise for time series processes with bounded support. We obtain that the restricted PVQMLEs retain the  desired asymptotic properties when the imposed  restrictions are valid with respect to the true parameter of the mean and a pseudo-true parameter of the conditional variance. The validity of such restrictions can be tested without requiring correct specification of the conditional variance. We  derive a test for this purpose that can be used as a consistency test for restricted PVQMLEs. When the evidence-based parameter constraints are  identified and validated, they constitute a restriction set where an higher-efficiency restricted PVQMLE  can be obtained. Furthermore, under correct specification of the pseudo-variance,  the test can be used as a specification test on the underlying process generating the data.

Finally, the practical usefulness of PVQMLE approach is illustrated by means of two real data applications. One is concerned with  INAR models and one with a Beta autoregression for double-bounded data. INAR processes depend  on the distribution assumed for the innovation  and the thinning specification \citep{lu2021predictive}. \cite{guerrero2022integer} consider  an alternative INAR parametrization that is based on the innovation and marginal distributions that leads to an equivalent INAR specification where the thinning operator is specified implicitly. Our test allows us to test for the degree of dispersion in the thinning operator as well as the error term.  There exists a vast literature of INAR models in testing innovations and marginal distributions dispersion \cite[]{schweer2014compound, aleksandrov2020testing}, testing for serial dependence \cite[]{sun2013score}, and general goodness of fit tests \cite[]{weiss2018goodness}. However,  to the best of our knowledge, specification  tests are not available  for the thinning dispersion.  The thinning operator is typically  assumed to be binomial, which implies underdispersion in the thinning.  
Once appropriate thinning and innovation restrictions are identified through the specification test, the  corresponding PVQMLE is used to estimate the parameters of the INAR model. The second application concerns the analysis of daily realized correlations between a pair of stock returns, which forms a double-bounded time series as  the realized correlation takes values between minus one and one. We consider a pseudo-variance specification based on the implied variance from  Beta-distributed variables for the definition of PVQMLEs. We then test the validity of parametric restrictions between the mean and pseudo-variance to validate  the use of restricted PVQMLEs.

The remainder of the paper is organized as follows. Section~\ref{SUBSEC: Model} introduces the general mean and pseudo-variance framework and the  PVQMLEs,  together with some  examples. Section~\ref{SEC: Asymp. theory} presents the main theoretical results of the paper on the asymptotic properties of the PVQMLE and some special cases. Section~\ref{SEC: Efficiency} discusses the efficiency of the PVQMLE.
 Section~\ref{SEC: Specification tests} introduces the specification test for the validity of the constraints  with an extensive simulation study in the case of INAR models. 
Section~\ref{SEC: Applications} presents empirical applications.   Section~\ref{SEC: conclusions} concludes the paper. The proofs of the main results are deferred to Appendix~\ref{Appendix A proof of results}. Finally, Appendix~\ref{Appendix B further numerical results} includes additional numerical results.

\section{Specification and estimation} 
\label{SUBSEC: Model}
\subsection{PVQML estimators} 
Consider a stationary and ergodic time series process  $\{Y_t\}_{t\in\mathbb{Z}}$ with elements taking values in the sample space $\mathcal{Y}\subseteq \mathbb{R}$ and   with conditional mean given by
\begin{equation}
	\E(Y_t|\Fb_{t-1})= \lambda(Y_{t-1}, Y_{t-2}, \dots; \psi_0)=\lambda_t(\psi_0)\,, \quad t\in\Z,
	\label{mean equation}
\end{equation}
where $\Fb_t$ denotes the $\sigma$-field generated by $\{Y_{s}\,,\,\, s\leq t\}$, $ \lambda : \R^\infty \times \Psi \to \R$ is a known measurable function, and $\psi_0\in \Psi \subset \R^{p}$ is the true unknown  $p$-dimensional parameter vector.  We denote with $\nu_t$ the conditional variance of the process, i.e.~$\V(Y_t|\Fb_{t-1})  = \nu_t$, which is considered to have an unknown specification. The model is a semi-parametric model as the quantity of interest is the parameter vector of the conditional mean $\psi_0$   and other distributional properties are left unspecified and treated as an infinite dimensional nuisance parameter. The general specification of the model in \eqref{mean equation} includes a wide range of time series models as special case. For instance, it includes  linear and non-linear ARMA models  when $\mathcal{Y}=\mathbb{R}$, INGARCH and INAR models when $\mathcal{Y}=\mathbb{N}$, ACD models when $\mathcal{Y}={(0,\infty)}$, and Beta autoregressive models for bounded data when $\mathcal{Y}=(0,1)$. 

The main objective is to estimate the parameter vector $\psi_0$ of the conditional expectation. For this purpose, we  consider the specification of a   pseudo-variance 
\begin{equation}
	\nu^*_t(\gamma)= \nu^*(Y_{t-1}, Y_{t-2}, \dots; \gamma), \quad t\in\Z,
	\label{pseudo var equation}
\end{equation}
where $ \nu^* : \R^\infty \times \Gamma \to \left[ 0, +\infty \right) $ is a known  function that is indexed by the $k$-dimensional parameter $\gamma\in\Gamma\subset  \mathbb{R}^k$. 
We refer to this as a pseudo-variance as it is not necessarily correctly specified, i.e.~there may be no value $\gamma\in\Gamma$ such that $\nu^*_t(\gamma)=\nu_t$. The idea is to use the pseudo-variance $\nu^*_t(\gamma)$ to enhance the efficiency of the estimation of $\psi_0$ by means of a Gaussian QMLE. We denote the whole parameter vector that contains both the parameter of the mean and pseudo-variance with $\theta=(\psi',\gamma')'$ and $\theta\in \Theta=\Psi \times \Gamma \subset \R^m$, $m=p+k$.

We introduce the class of PVQMLEs that relies  on a Gaussian quasi-likelihood for the mean equation with the pseudo-variance as scale of the Gaussian density. We consider estimators based on both unrestricted and restricted quasi-likelihood functions. Assume that we have an observed sample of size $T$ from the process defined in \eqref{mean equation}, given by $\{Y_t\}_{t=1}^T$. Since $\lambda_t(\psi)$ and $\nu^*_t(\gamma)$ can depend on the infinite past of $Y_t$, we  define their  approximations of $\tilde \lambda_t(\psi)$ and $\tilde \nu^*_t(\gamma)$ based on the available finite sample $\{Y_t\}_{t=1}^T$,
\begin{equation}
	\tilde \lambda_t(\psi) = \lambda(Y_{t-1}, \dots, Y_1, \tilde Y_0, \tilde Y_{-1}, \dots; \psi)\,, \,\,\,\,\, 
	\tilde \nu^*_t(\gamma) = \nu^*(Y_{t-1}, \dots, Y_1, \tilde Y_0, \tilde Y_{-1}, \dots; \gamma),
	\label{proxy uncons}
\end{equation}
where $\tilde Y_0, \tilde Y_{-1}, \dots $ are given initial values. The Gaussian quasi-likelihood for $\psi$ with the pseudo-variance scaling  is defined as
\begin{equation}
	\tilde L_T(\theta)=\frac{1}{T}\sum_{t=1}^{T} \tilde l_t(\theta)\,,   \quad \tilde l_t(\theta)=-\frac{1}{2} \log \tilde \nu^*_t(\gamma) -\frac{[Y_t-\tilde \lambda_t(\psi)]^2}{2 \tilde \nu^*_t(\gamma)}.
	\label{quasi likelihood}
\end{equation}

Based on the quasi-likelihood function in \eqref{quasi likelihood}, we define the  unrestricted and restricted PVQMLE. The unrestricted PVQMLE is based on the unconstrained maximization of the pseudo-likelihood without imposing any constrains between  $\psi$ and  $\gamma$. The  unrestricted  PVQMLE $\hat\theta$ is defined as
\begin{equation}
	\hat \theta = \argmax_{\theta \in \Theta } \tilde L_T(\theta),
	\label{unconstrained}
\end{equation}
where $\hat \theta=(\hat \psi', \hat \gamma')'$ and $\hat \psi$ is the unrestricted PVQMLE of $\psi_0$. In Section \ref{SEC: Asymp. theory}, we shall see that the unrestricted PVQMLE $\hat \psi$ is a consistent estimator of $\psi_0$ and, in fact, it is asymptotically equivalent to a specific WLSE. If the pseudo-variance is correctly specified, i.e.~there is $\gamma_0\in \Gamma$ such that $\nu^*_t(\gamma_0)=\nu_t$, then $\hat \psi$ is asymptotically equivalent to the efficient WLSE.

In models where the sample space $\mathcal{Y}$ is bounded, such as count-time series models, there can be a natural relationship between the conditional mean and variance of the process. For example, in a count time series process we have that if the mean goes to zero, then also the variance goes to zero as, in fact, the limit case is the mean being exactly zero.  Such relationship between mean and variance, as given by parametric models, provide  a natural way to introduce restrictions between the mean and pseudo-variance parameters $\psi$ and $\gamma$. Several examples are presented at the end of this section.

To specify the restricted PVQMLE, we consider  the constrained parameter set $\Theta_R$ that imposes $r$ restrictions on the pseudo-variance parameters 
$$\Theta_R=\{\theta\in\Theta: S\gamma=g(\psi)\},$$ 
where $S$ is a $r\times k$ selection matrix and $g : \Psi \to \mathbb{R}^r$. The estimator derived from the maximization of \eqref{quasi likelihood} over the set $\Theta_R$ is the restricted PVQMLE,
\begin{equation}
	\hat \theta_R = \argmax_{\theta \in \Theta_R} \tilde L_T(\theta)
	\label{constrained}
\end{equation}
where $\hat \theta_R=(\hat \psi_R', \hat \gamma_R')'$ and $\hat \psi_R$ is the restricted PVQMLE of $\psi_0$. In Section \ref{SEC: Asymp. theory}, we shall see that the  restricted PVQMLE $\hat \psi_R$ is a consistent estimator of $\psi_0$ if the  constrains in $\Theta_R$ hold with respect to a pseudo-true parameter $\gamma^*$. The advantage of the  restricted PVQMLE $\hat \psi_R$  is that   it can achieve higher efficiency than the unrestricted one. Furthermore, as it shall be  presented in Section \ref{SEC: Specification tests}, the validity of the restrictions can be tested under both misspecification and correct specification of the pseudo-variance. The test can be interpreted as a consistency test for the restricted estimator when the pseudo-variance is misspecified. Instead, it can be employed as  a specification test if we assume correct specification of the pseudo-variance. For instance, it shall be employed to test for underdispersion, equidispersion or overdispersion in the thinning operator of INAR models.

In practice, both the unrestricted and the restricted PVQMLE \eqref{unconstrained}-\eqref{constrained} do not have a closed form solution. Therefore, the estimation of model parameters is carried out by numerical optimization. This is done by employing standard optimization functions of the \texttt{R} software using the BFGS algorithm \citep{nocedal1999numerical}. 

\subsection{Examples}

The model specification in \eqref{mean equation}  is very general and it covers a wide range of semi-parametric observation-driven time series model. The unrestricted and restricted  QMLE based on the pseudo-variance in \eqref{pseudo var equation}  can be employed for such general class of models. However,  PVQMLEs are particularly suited for time series  processes where the support of the conditional mean is bounded and a natural relationship with the conditional variance can be assumed. In Section \ref{SUBSEC: Improved efficiency}  it will be shown that in models where conditional mean and pseudo-variance share some parameter restrictions, a more efficient estimator may be obtained with respect to alternative estimation approaches available in the literature.
The specification of the pseudo-variance and the parameter restrictions with the conditional mean can be based on well known model specifications. The validity of such restrictions is testable  and the asymptotic properties do not require correct specification of the pseudo-variance. This means that no assumptions on the true conditional variance are needed and the consistency of the restricted PVQMLE can also be tested  without relying on such assumptions. Below we  present  some examples of models that  are encompassed in the framework defined in equations \eqref{mean equation} and \eqref{pseudo var equation}, and provide a general way to specify the pseudo-variance and the parameter restrictions with the conditional mean.

\begin{ex}[INAR models] \rm
	\label{SUBSEC: INAR}
	
	INAR models   are widely used in the literature to model count time series.   The INAR($1$) model is given by
	\begin{align}\label{inar}
		Y_t =a \circ Y_{t-1} +\varepsilon_t \,, \quad t \in \Z,
	\end{align}
	where  $\{\varepsilon_t\}_{t\in \mathbb{Z}}$ is an iid sequence of non-negative integer-valued random variables with   mean $\omega_1 > 0$ and variance $\omega_2 > 0$, and  `$\circ$' is the thinning operator  of \cite{steutel1979discrete}. 
	For a given $N \in \N$, a general formulation of  the thinning operator is $a \circ N = \sum_{j=1}^{N} X_j$ when $N>0$, and 0 otherwise, where $X_j$ is a sequence of iid non-negative integer-valued random variables following a distribution with finite mean $a$ and variance $b$, say $X_j \sim D_X(a,b)$.
	The most common formulation \citep{steutel1979discrete} is the binomial thinning where  $X_j$ is a sequence of  independent Bernoulli random variables with success probability $a \in (0,1)$, therefore $a \circ N$ is a binomial random variable with $N$ trials and success probability $a$. 
	The conditional mean of the INAR($1$) is  
	\begin{equation} \label{inar mean}
			\lambda_t =a  Y_{t-1}+ \omega_1\,.
	\end{equation}
	The form of the variance for INAR models is known to be linear in the observations, therefore the pseudo-variance can be specified as
	\begin{equation} \label{inar pseudo-var}
	\nu_t^*=b Y_{t-1}+\omega_2\,.
	\end{equation}
	Several restrictions can be considered for the PVQMLE. For instance, the restriction $b=a(1-a)$ is implied by a binomial thinning and  $\omega_1=\omega_2$ is implied by a Poisson error. 
The same estimation framework applies to INAR models with general lag order $p$, called INAR($p$).
\begin{equation*} 
	Y_t =a_1 \circ Y_{t-1}+ \dots +a_p \circ Y_{t-p} +\varepsilon_t \,, \quad t \in \Z,
\end{equation*}
\begin{equation*}
	\lambda_t= \sum_{h=1}^{p}a_h  Y_{t-h} + \omega_1\,, ~~~~~~
	\nu_t^*=\sum_{h=1}^{p}b_h Y_{t-h}+\omega_2.
\end{equation*}
Further results on INAR models are discussed in Section~\ref{SEC: inar} and \ref{SEC: Specification tests}. An application to real data is presented in Section~\ref{SEC: Applications}.
\end{ex}

\begin{ex}[INGARCH models] \rm 
	Another popular model for time series of counts is the INGARCH model. The conditional mean of the  INGARCH($1$,$1$) model takes the form
	\begin{equation}\label{ingarch}
		\lambda_t= \omega_1 + \alpha_1 Y_{t-1} + \beta_1  \lambda_{t-1}\,,
	\end{equation}
	where $\omega_1,\alpha_1,\beta_1  \ge 0$. The pseudo-variance can be specified as
	\begin{equation} \label{ingarch pseudo-var}
	\nu_t^*= \omega_2 + \alpha_2 Y_{t-1} + \beta_2  \lambda_{t-1}\,.
	\end{equation}
	Also in this case, several restrictions can be considered for the PVQMLE. For instance, the restrictions $\omega_2=\omega_1$, $\alpha_2=\alpha_1$ and $\beta_2=\beta_1$ are implied by an equidispersion assumption $\nu_t^*=\lambda_t$, which follows assuming a conditional Poisson distribution for example. Alternatively, the restrictions $\omega_2=c\omega_1$, $\alpha_2=c\alpha_1$ and $\beta_2=c\beta_1$ with $c>0$ are implied by a proportional variance assumption $\nu_t^*=c\lambda_t$. 
	

\end{ex}

\begin{ex}[ACD models] \rm 
	ACD  models   are typically used  to model non-negative continuous time series variables, like durations or volumes. These models take the form $Y_t=\lambda_t\varepsilon_t$ where $\varepsilon_t$ is a sequence of positive variables with mean equal to 1. The conditional expectation $\lambda_t$ may take the form as in equation \eqref{ingarch}.  The pseudo-variance can be specified in several ways and restrictions can be imposed. For instance, the restriction  $\nu_t^*=\lambda_t^2$ follows by assuming an exponential error distribution. An alternative restriction is given by $\nu_t^*=c\lambda_t^2$, $c>0$.

	
\end{ex}

\begin{ex}[double-bounded autoregressions] \rm \label{Ex. beta}
	For double-bounded time series data  the conditional mean $\lambda_{t}$ can be specified  as in equation \eqref{ingarch}, see \cite{gorgi2021beta} for instance. Several specifications and restrictions for the pseudo-variance can be considered. For instance, the 
	restriction $\nu^*_t= \lambda_{t}(1-\lambda_{t})/(1+\phi)$ is implied by a beta conditional distribution with dispersion parameter $\phi>0$. Intermediate  restrictions on the pseudo-variance   are discussed in the corresponding application in Section \ref{SEC: Applications}. See also Section~\ref{SEC: beta} for further results established on this class of models.
\end{ex}

We note that the examples presented in this section are focused on a linear mean equation for simplicity of exposition. Several other non-linear model specifications are encompassed in the general framework in \eqref{mean equation} and \eqref{pseudo var equation}, see for example \cite{Creal2013} and \cite{christou_fokianos_2015}.

\section{Asymptotic theory} 
\label{SEC: Asymp. theory}
In this section, the asymptotic properties of the PVQMLEs in \eqref{unconstrained} and \eqref{constrained} are formally derived. Although asymptotic results related to quasi-maximum likelihood estimators of observation-driven models are well-established in the literature,  the associated theory for PVQMLEs differs as it relies on simultaneous estimation of mean and pseudo-variance parameters, where the latter can be misspecified and present parameter restrictions with the mean. Since  the pseudo-variance can be misspecified, the estimator of the pseudo-variance parameter $\hat\gamma$ will be consistent with respect to a pseudo-true value $\gamma^*$, which is given by
\begin{equation} \label{pseudo_true_par}
	\gamma^* = \argmax_{\gamma\in \Gamma} -\frac{1}{2}\E\left( \log \nu^*_t(\gamma) + \frac{[Y_t- \lambda_t(\psi_0)]^2}{  \nu^*_t(\gamma)} \right).
\end{equation}
We  define the vector $\theta_0 = (\psi_0^\prime,\gamma^{*\prime})^\prime$ that contains both true and pseudo-true parameters. The estimator of the mean parameters preserves the  consistency and asymptotic normality results to the true parameter vector $\psi_0$ irrespective of the correct specification of the conditional variance. We  show that such result holds for both unrestricted \eqref{unconstrained} and restricted \eqref{constrained} estimators, where the restricted estimator requires the validity of the imposed restrictions with respect to the pseudo-true parameter, i.e.~$S\gamma^*=g(\psi_0)$. We note that the validity of such restriction is a weaker condition than the correct specification of the pseudo-variance. In fact, the  test proposed in Section \ref{SEC: Specification tests} is a restriction test and, under the null hypothesis of valid restrictions, the pseudo-variance can still be misspecified. 

We start by showing consistency and asymptotic normality of the unrestricted PVQMLE in \eqref{unconstrained}. We first obtain the  score function related to \eqref{quasi likelihood} 
\begin{equation} \label{score}
	\tilde S_T(\theta)=\frac{1}{T}\sum_{t=1}^{T} \tilde s_t(\theta), ~~\, \tilde s_t(\theta)=\frac{Y_t- \tilde \lambda_t(\psi)}{\tilde \nu^*_t(\gamma)} \frac{\partial \tilde \lambda_t(\psi)}{\partial \theta} + \frac{[Y_t- \tilde \lambda_t(\psi)]^2-\tilde \nu^*_t(\gamma)}{2 \tilde \nu^{*2}_t(\gamma)}  \frac{\partial \tilde \nu^*_t(\gamma)}{\partial \theta}.
\end{equation}
Then, define $L_T(\theta)$, $l_t(\theta)$,  $S_T(\theta)$ and $s_t(\theta)$ as the random functions obtained from $\tilde{L}_T(\theta)$,  $\tilde{l}_t(\theta)$, $\tilde{S}_T(\theta)$ and  $\tilde{s}_t(\theta)$ by substituting $\tilde{\lambda}_t(\psi)$ and  $\tilde \nu^*_t(\gamma)$  with $\lambda_t(\psi)$ and  $\nu^*_t(\gamma)$, respectively. Furthermore,
let $H(\theta_0) = \E [ - \partial^2 l_t(\theta_0)/\partial \theta \partial \theta^{\prime} ] $ and $I(\theta_0)= \E[s_t(\theta_0)s_t(\theta_0)^\prime ]$.
Consider the following assumptions.

\begin{enumerate}[label=\textbf{A\arabic*}]
	\item The process $\{ Y_t, \lambda_t \}_{ t \in Z }$  is strictly stationary and ergodic. 
	\label{Ass. stationarity}
	\item \label{Ass. uniform moment} 
	$\lambda_t(\cdot)$ is continuous in $\Psi$, $\nu^*_t(\cdot)$ is continuous in $\Gamma$ and the set $\Theta$ is compact. Moreover,
	\begin{equation*}
		\E \,\, \sup_{\gamma \in \Gamma} \norm{\log \nu^*_t(\gamma)} < \infty\,, \quad \E \,\, \sup_{\theta \in \Theta} \frac{[Y_t- \lambda_t(\psi)]^2}{\nu^*_t(\gamma)}    < \infty\,.
	\end{equation*}
	\item \label{Ass. identification} $\lambda_t(\psi) = \lambda_t(\psi_0)$ a.s. if and only if $\psi=\psi_0$. 
	\item There is a constant $\underline{\nu}^* > 0$ such that $\nu^*_t(\gamma), \tilde\nu^*_t(\gamma)  \geq \underline{\nu}^*$  for any $t \geq 1$ and any $\gamma \in \Gamma$. \label{Ass. lower bound}
	\item \label{Ass. starting value}
	Define  $a_t = \sup_{\psi \in \Psi} |\tilde \lambda_t(\psi)-\lambda_t(\psi)| $ and $b_t = \sup_{\gamma \in \Gamma} \norm{\tilde \nu^*_t(\gamma)-\nu^*_t(\gamma)} $, it holds that
	$$
	\lim_{t \to \infty} \Big( 1 + \norm{Y_t} + \sup_{\psi \in \Psi} |\lambda_t(\psi)|\Big)  a_t = 0\,, \quad \lim_{t \to \infty} \Big( 1 + Y^2_t + \sup_{\psi \in \Psi} \lambda^2_t(\psi)\Big)  b_t = 0 \quad a.s.
	$$
	
	\item The pseudo-true parameter $\gamma^* \in \Gamma$  defined in \eqref{pseudo_true_par} is unique.
	\label{Ass. unique maximizer}
	
	\item \label{Ass. starting value derivative}
	Define $ c_t = \sup_{\theta \in \Theta} \|\partial \tilde \lambda_t(\psi)/\partial \theta- \partial\lambda_t(\psi)/\partial \theta\| $, $ d_t = \sup_{\theta \in \Theta} \lnorm{\partial \tilde \nu^*_t(\gamma)/\partial \theta- \partial \nu^*_t(\gamma)/\partial \theta} $. The following quantities are   of order $\Ob(t^{-\delta})$ a.s. for some $\delta > 1/2$ 
	$$ \sup_{\theta \in \Theta}\lnorm{\frac{\partial\lambda_t(\psi)}{\partial \theta}} a_t \,, \quad \sup_{\theta \in \Theta}\lnorm{\frac{\partial\nu^*_t(\gamma)}{\partial \theta}}\Big(  1 + \norm{Y_t} + \sup_{\psi \in \Psi} \norm{\lambda_t(\psi)} \Big)  a_t \,,  $$
	$$ \sup_{\theta \in \Theta}\lnorm{\frac{\partial\lambda_t(\psi)}{\partial \theta}}\Big(  \norm{Y_t} + \sup_{\psi \in \Psi} \norm{\lambda_t(\psi)} \Big)  b_t\,, \quad  \sup_{\theta \in \Theta}\lnorm{\frac{\partial\nu^*_t(\gamma)}{\partial \theta}} \Big( 1 + Y^2_t+\sup_{\psi \in \Psi} \lambda_t^2(\psi)\Big) b_t \,,  $$
	$$ \big(  1 + \norm{Y_t} + \sup_{\psi \in \Psi} \norm{\lambda_t(\psi)} \big)  c_t \,, \quad \big( 1 + Y^2_t+\sup_{\psi \in \Psi} \lambda_t^2(\psi)\big) d_t.~~~~~~$$
	\item \label{Ass. moment second derivative}
	$\lambda_t(\cdot)$ and $\nu^*_t(\cdot)$ have continuous second-order derivatives in their spaces. Moreover, 
	$$
	\E \,\, \sup_{\theta \in \Theta} \frac{[Y_t- \lambda_t(\psi)]^4}{\nu^{*2}_t(\gamma) }   < \infty\,, \quad \E \,\, \sup_{\theta \in \Theta}  \lnorm{\frac{1}{\sqrt{\nu^{*}_t(\gamma)}}\frac{\partial^2\lambda_t(\psi)}{\partial \theta \partial \theta^\prime}}^2  < \infty\,,
	$$
	$$
	\E \,\, \sup_{\theta \in \Theta} \lnorm{\frac{1}{\nu^*_t(\gamma)}\frac{\partial\lambda_t(\psi)}{\partial \theta } \frac{\partial\lambda_t(\psi)}{\partial \theta^\prime }}  < \infty\,, \quad \E \,\, \sup_{\theta \in \Theta} \lnorm{\frac{1}{\nu^*_t(\gamma)}\frac{\partial\lambda_t(\psi)}{\partial \theta } \frac{\partial\nu^*_t(\gamma)}{\partial \theta^\prime }}^2  < \infty\,,
	$$
	$$
	\E \,\, \sup_{\theta \in \Theta} \lnorm{\frac{1}{\nu^{*2}_t(\gamma)}\frac{\partial\nu^*_t(\gamma)}{\partial \theta } \frac{\partial\nu^*_t(\gamma)}{\partial \theta^\prime }}^2  < \infty\,, \quad \E \,\, \sup_{\theta \in \Theta} \lnorm{\frac{1}{\nu^*_t(\gamma)}\frac{\partial^2\nu^*_t(\gamma)}{\partial \theta \partial \theta^\prime }}^2  < \infty\,,
	$$
	$$
	\E \,\, \frac{[Y_t- \lambda_t(\psi_0)]^8}{\nu^{*4}_t(\gamma^*) }   < \infty\,, \quad \E \,\, \lnorm{\frac{1}{\nu^*_t(\gamma^*)}\frac{\partial\lambda_t(\psi_0)}{\partial \theta } \frac{\partial\lambda_t(\psi_0)}{\partial \theta^\prime }}^2  < \infty\,.
	$$
	

	\item \label{Ass. hessians}  
	The matrices $H(\theta_0)$ and $I(\theta_0)$ are positive definite.
	
	\item $\theta_0 \in \dot{\Theta}$, where $\dot{\Theta}$ is the interior of $\Theta$. \label{Ass. interior}
	
	\item The sequence $\sqrt{T}S_T(\theta_0)$ obeys the central limit theorem. \label{Ass. score clt}
	
\end{enumerate}

The strict stationarity and ergodicity   in assumption \ref{Ass. stationarity}   depends upon the model formulation in \eqref{mean equation} and \eqref{pseudo var equation} and it can  be established by means of different probabilistic approaches, see for instance \cite{straumann2006} and \cite{tru2021}. Assumption \ref{Ass. uniform moment} is  a standard moment condition.
Assumption \ref{Ass. identification} is  required for the identification of the true parameter $\psi_0$. Assumptions \ref{Ass. starting value} and \ref{Ass. starting value derivative} are needed to guarantee that the  initialization of filters in \eqref{proxy uncons} is asymptotically irrelevant.  Assumption \ref{Ass. unique maximizer} imposes the uniqueness of the pseudo-true parameter for the variance equation. In Corollary~\ref{Cor. set consistency} below, we show that this assumption can be dropped if  the researcher is not interested in the asymptotic normality of the estimator but only in the consistency. 
Assumption \ref{Ass. moment second derivative} imposes moments on the second derivatives of the log-quasi-likelihood that are required for asymptotic normality to apply. 
Assumption \ref{Ass. hessians} is required to obtain the positive definiteness of the asymptotic covariance matrix of the estimators. This condition is left high-level for generality purposes. However, in Lemma~\ref{Lem invertibility} in appendix~\ref{Appendix subsec technical}, we introduce some special cases and sufficient low-level conditions that verify the assumption.
 Assumption \ref{Ass. interior} is the standard condition for asymptotic normality that the pseudo-true parameter value is in the interior of the parameter set. Finally, assumption \ref{Ass. score clt} is an high-level condition that a central limit theorem applies to the score. This condition is also left high-level for generality purposes  since the score function $s_t(\theta_0)$  is not a martingale difference sequence, see equation \eqref{score}.  There are several alternative Central Limit Theorems (CLT) for non-martingale sequences and the choice of the most appropriate one is strongly dependent on the specific mean-variance model formulation. For example, CLTs appealing the concept of mixing processes or mixingales are widely available, see the surveys in \cite{douk1994}, \cite{bradley_2005} and \cite{white1994estimation}. See also the proof of Theorem~\ref{Thm. can inar} below for an example in which the assumption is satisfied by appealing the CLT for $\alpha$-mixing processes.
Finally, in case of correct conditional variance specification then assumption \ref{Ass. score clt} can be dropped, see Corollary~\ref{Cor. correct variance}. Theorem  \ref{Thm. can} delivers the consistency and asymptotic normality of the unrestricted PVQMLE of the true parameter $\psi_0$.

\begin{theorem} \label{Thm. can}
	Consider the unrestricted PVQMLE in  \eqref{unconstrained}. Under conditions~\ref{Ass. stationarity}-\ref{Ass. unique maximizer}
	\begin{equation} \label{consistency}
		\hat{\psi} \xrightarrow{} \psi_0\,, \quad a.s. \quad T \to \infty\,.
	\end{equation}
	Moreover, if also \ref{Ass. starting value derivative}-\ref{Ass. score clt} hold, as $T \to \infty$
	\begin{equation} \label{asymp. normality}
		\sqrt{T} \left( \hat{\psi} - \psi_0 \right)  \xrightarrow{d} N(0, \Sigma_\psi)\,, \quad \quad \Sigma_\psi=H_\psi^{-1}(\theta_0) I_\psi(\theta_0) H_\psi^{-1}(\theta_0)\,,
	\end{equation}
	where 
	\begin{equation}
		H_\psi(\theta_0) = 
		\E\left[ 
		\frac{1}{\nu^*_t(\gamma^*)} \frac{\partial \lambda_t(\psi_0)}{\partial \psi} \frac{\partial \lambda_t(\psi_0)}{\partial \psi^\prime} 
		\right], \,\,
		I_\psi(\theta_0)= 
		\E\left[ 
		\frac{\nu_t}{\nu^{*2}_t(\gamma^*)} \frac{\partial \lambda_t(\psi_0)}{\partial \psi} \frac{\partial \lambda_t(\psi_0)}{\partial \psi^\prime}
		\right]. \label{hessians uncons}
	\end{equation}
	In addition, $\Sigma_\psi$ is positive definite.
\end{theorem}




The asymptotic  properties of the estimator of the pseudo-variance parameters $\gamma$ are obtained from Theorem~\ref{Thm. can} as a byproduct. We make the result explicit  in Corollary \ref{Cor. can gamma} below. Let $s_t(\theta_0)=[s_t^{(\psi)}(\theta_0)^\prime, s_t^{(\gamma)}(\theta_0)^\prime ]^\prime $ be the partition of the score with respect to the mean and (pseudo-)variance parameters.  Define the partitions $H_{\gamma}(\theta_0) = \E [ - \partial^2 l_t(\theta_0)/\partial \gamma \partial \gamma^{\prime} ] $ and $I_\gamma(\theta_0)= \E[s_t^{(\gamma)}(\theta_0)s_t^{(\gamma)}(\theta_0)^\prime ]$.

\begin{corollary} \label{Cor. can gamma}
	Under the assumptions of Theorem~\ref{Thm. can} we have that as $T \to \infty $, a.s.  $\hat{\gamma} \xrightarrow{} \gamma^*$ and $\sqrt{T} \left( \hat{\gamma} - \gamma^* \right)  \xrightarrow{d} N(0, \Sigma_\gamma)$, where $ \Sigma_\gamma=H_\gamma^{-1}(\theta_0) I_\gamma(\theta_0) H_\gamma^{-1}(\theta_0)$. In addition, $\Sigma_\gamma$ is positive definite.
\end{corollary}

Theorem~\ref{Thm. can} determines  the asymptotic distribution of the unrestricted PVQMLE of $\psi_0$ without requiring correct specification of the pseudo-variance. The following result shows that in the special case in which the variance is well-specified then the  estimator $\hat{\psi}$ gains in efficiency.
\begin{corollary} \label{Cor. correct variance}
	Consider the assumptions of Theorem~\ref{Thm. can}. If, in addition, the variance \eqref{pseudo var equation} is correctly specified, i.e.~$\nu^*_t(\gamma^*) = \nu_t$, then \ref{Ass. stationarity}-\ref{Ass. interior} entail \eqref{consistency} and
	\begin{equation}
		\sqrt{T} \left( \hat{\psi} - \psi_0 \right)  \xrightarrow{d} N(0, I^{-1}_\psi)\,, \quad \quad I_\psi 
		= \E\left[ 
		\frac{1}{\nu_t} \frac{\partial \lambda_t(\psi_0)}{\partial \psi} \frac{\partial \lambda_t(\psi_0)}{\partial \psi^\prime} 
		\right]\,,
		\label{hessian wlse}
	\end{equation}
	where $\Sigma_\psi - I_\psi^{-1}$ is positive semi-definite.
\end{corollary}

We also note that in Corollary \ref{Cor. correct variance} the uniqueness   of the variance parameter  in assumption \ref{Ass. unique maximizer} is implied by the condition $\nu^*_t(\gamma) = \nu^*_t(\gamma^*) $ a.s.~if and only if $\gamma=\gamma^*$. This follows immediately from the correct specification of the pseudo-variance.
Corollary \ref{Cor. set consistency} below shows that even if the pseudo-true parameter  $\gamma^*$ is not unique, i.e.~assumption  \ref{Ass. unique maximizer} does not hold, the consistency of the unrestricted estimator  $\hat \psi$ is retained without any additional assumption. The overall estimator $\hat \theta$ will instead be set consistent over the set of values that maximize  the limit of the quasi-likelihood, $\Theta_0$, since the pseudo-true parameter $\gamma^*$ is not uniquely identified. 
%
%
\begin{corollary} \label{Cor. set consistency}
	Consider the unrestricted PVQMLE  \eqref{unconstrained} and assume conditions \ref{Ass. stationarity}-\ref{Ass. starting value} 
	hold.  Then, as $T \to \infty $, $\inf_{\theta_0 \in \Theta_0} \|{ \hat{\theta} - \theta_0 }\| \xrightarrow{} 0$ a.s.~and  $\hat{\psi} \xrightarrow{} \psi_0$  a.s.
\end{corollary}


We now treat the case in which the conditional mean and  pseudo-variance parameters are constrained. We study the asymptotic properties of the restricted PVQMLE $\hat \psi_R$  defined in \eqref{constrained}.  

\begin{enumerate}[label=\textbf{A12}] 
	\item \label{Ass. constrain holds} The equality $S \gamma^*=g(\psi_0)$ holds and $g(\cdot)$ is continuous. 
\end{enumerate}	

Assumption \ref{Ass. constrain holds} is required to   ensure that $\theta_0 \in \Theta_R$, i.e.~the imposed restrictions are valid with respect to the true parameter $\psi_0$ and the pseudo-true parameter $\gamma^*$. The continuity of $g(\cdot)$ guarantees that $\Theta_R$ remains compact. Define $\gamma=(\gamma'_1, \gamma'_2)'$ where $\gamma_1=S\gamma=g(\psi)$ is the sub-vector of pseudo-variance parameters that are restricted to  mean parameters and $\gamma_2$ constitutes the sub-vector of remaining free parameters. For $\theta \in \Theta_R$, we have $\theta=(\psi',\gamma'_1,\gamma'_2)'=(\psi',g(\psi)',\gamma'_2)'$ so the $m$-dimensional vector of parameters to estimate is reduced to $\theta = (\psi',\gamma'_2)'$, with some  abuse of notation. The new parameter vector has dimension $m_R=p+k_2$ where $k_2$ is the length of the extra nuisance parameters $\gamma_2$. Recall that $H_x(\theta_0) = \E \left[ - \partial^2 l_t(\theta_0)/\partial x \partial x^{\prime} \right] $ and $I_x(\theta_0)= \E [s_t^{(x)}(\theta_0)s_t^{(x)}(\theta_0)^\prime ]$. Moreover, define $H_{x,z}(\theta_0) = \E \left[ - \partial^2 l_t(\theta_0)/\partial x \partial y^{\prime} \right] $, $I_{x,z}(\theta_0)= \E [s_t^{(x)}(\theta_0)s_t^{(z)}(\theta_0)^\prime ]$ and $I_{z,x}(\theta_0)=I'_{x,z}(\theta_0)$. Analogously, set $D(\theta_0)=H^{-1}(\theta_0)$ and $D_{x,y}(\theta_0)$ being the corresponding partition related to rows $x$ and columns $y$ of $D(\theta_0)$. Theorem \ref{Thm. constrained can} delivers the asymptotic distribution of the restricted PVQMLE.
\begin{theorem} \label{Thm. constrained can}
	Consider the restricted PVQMLE in  \eqref{constrained}. Under conditions~\ref{Ass. stationarity}-\ref{Ass. unique maximizer} and \ref{Ass. constrain holds}
	\begin{equation} \label{consistency cons}
		\hat{\psi}_R \xrightarrow{} \psi_0\,, \quad a.s. \quad T \to \infty\,.
	\end{equation}
	Moreover, if also \ref{Ass. starting value derivative}-\ref{Ass. score clt} hold, as $T \to \infty$
	\begin{equation} \label{asymp. normality cons}
		\sqrt{T} \left( \hat{\psi}_R - \psi_0 \right)  \xrightarrow{d} N(0, \Sigma_{R}) \,,
	\end{equation}
	where 
	\begin{align} \label{covariance constr}
		\Sigma_{R}=& D_{\psi}(\theta_0) I_{\psi}(\theta_0) D_{\psi}(\theta_0)	+ D_{\psi,\gamma_2}(\theta_0) I_{\gamma_2,\psi}(\theta_0) D_{\psi}(\theta_0)		 \\
		&+ D_{\psi}(\theta_0) I_{\psi,\gamma_2}(\theta_0) D_{\gamma_2,\psi}(\theta_0)	
		+ D_{\psi,\gamma_2}(\theta_0) I_{\gamma_2}(\theta_0) D_{\gamma_2,\psi}(\theta_0)	\,. \nonumber
	\end{align}
	In addition, $\Sigma_R$ is positive definite.
\end{theorem}

We note that Corollaries~\ref{Cor. can gamma}-\ref{Cor. set consistency} can easily be adapted to hold also for $\hat \theta_R$. In Section~\ref{SUBSEC: Improved efficiency} below, we shall see  that the restricted PVQMLE    can lead to substantial gains in efficiency with respect to the unrestricted PVQMLE. The consistency of the  restricted PVQMLE requires the additional assumption \ref{Ass. constrain holds}. However, as discussed in Section  \ref{SEC: Specification tests}, this assumption can be tested and the correct specification of the pseudo-variance is not required. Clearly,  when $\psi$ and $\gamma$ do not have parameter restrictions, i.e.~$\hat \psi_R=\hat \psi$, it can be noted that Theorem  \ref{Thm. can} is equivalent to  Theorem \ref{Thm. constrained can} with $\Sigma_{R} = \Sigma_\psi$, since $H_{\psi,\gamma_2}(\theta_0)=0$, $H(\theta_0)$ becomes block diagonal, its inverse has block elements $D_x(\theta_0)=H_x^{-1}(\theta_0)$ and $D_{x,y}(\theta_0)=D_{y,x}(\theta_0)=0$, implying that $\Sigma_{R} = \Sigma_\psi$.

To illustrate the  relevance of the theoretical results, in the remainder of the section we provide  an application of the asymptotic results to two specific models of interest introduced in Section~\ref{SUBSEC: INAR}.

\subsection{Integer-valued autoregressive models} 
	\label{SEC: inar}
	
We consider the class of  INAR models specified in equation~\eqref{inar} with  the corresponding conditional mean given in \eqref{inar mean}.
Recall that for INAR models the thinning operator is defined as $a \circ N = \sum_{j=1}^{N} X_j$ when $N>0$, and 0 otherwise, where $X_j \sim D_X(a,b)$ are iid with finite mean $a$ and variance $b$. 
We start by studying the stochastic properties of the general class of INAR processes. Theorem \ref{Thm. stationarity inar} below provides  conditions   for strict stationarity and mixing properties of the INAR process. 
	\begin{theorem} \label{Thm. stationarity inar}
	Let the INAR process \eqref{inar} satisfy $a <1$. Then, the process admits
	a strictly stationary and ergodic solution with finite second moment $\E(Y_t^2)< \infty$. Moreover, the process is $\beta$-mixing with coefficients decaying geometrically fast.
	\end{theorem}
	Next, we derive the strong consistency
	and asymptotic normality of several PVQML estimators of INAR models by
	appealing to Theorems~\ref{Thm. can}-\ref{Thm. constrained can}. We assume that the observations are generated from an INAR(1) model with thinning and innovation distributions following some unspecified equidispersed distributions (i.e.~mean equal to the variance). We consider PVQMLEs for the parameter vector based on the  pseudo-variance  specified  in \eqref{inar pseudo-var}. We study the asymptotic properties of the unrestricted PVQMLE and the restricted PVQMLE with restrictions $\omega_2 =\omega_1$ and $b=a$. In this case, the restrictions hold but no  assumptions on the shape of the distribution of the data generating process are  imposed for the asymptotic results of the PVQMLE.  
		\begin{theorem} \label{Thm. can inar}
			Let $\{Y_1,\dots,Y_T\}$ be generated by the INAR(1) process  in \eqref{inar}  with  an equidispersed error, $\E(\varepsilon_t)=\V(\varepsilon_t)=\omega_{1}$, and an equidispersed thinning operator, $\E(a \circ N) =\V(a \circ N)=a N$, with $a<1$.
			Consider PVQMLEs for the parameter vector $\theta=(\omega_{1},a, \omega_{2},b)'$ based on the pseudo-variance $\nu^*_t$ specified   in \eqref{inar pseudo-var}. 
Furthermore, assume that $\theta_0 \in \Theta$, where $\Theta$ is a compact parameter set such that $\omega_1 >0$, $\omega_2 >0$, $a \geq 0$, $b \geq 0$. 
			Then, the unrestricted PVQMLE
			\eqref{unconstrained} and the following restricted PVQMLEs \eqref{constrained} with restrictions (i) $\omega_2 =\omega_1$, (ii)  $b=a$ and (iii) $(\omega_2 =\omega_1,b=a)$ are strongly consistent. 
			Assume further that $\theta_0 \in \dot{\Theta}$ and $\E(Y_t^8) < \infty$. 
			Then, all the PVQMLEs are also asymptotically normally distributed with asymptotic covariance matrix given in Theorems~\ref{Thm. can}-\ref{Thm. constrained can}. 
		\end{theorem}
	
		For the INAR model, the existence of the $r$-moments with $r>2$ depends on the specific discrete distribution for the errors $\varepsilon_t$ and the thinning. Since we keep such distributions unspecified, the existence of higher-order moments is required. However, the moment condition is satisfied for several INAR models. For example, when the thinning and the error distributions are Poisson
		the observation $Y_t$ are Poisson marginally distributed and then the moments of any order are finite \cite[Lem.~A.1]{chri2014}. For details on more general INAR modeling see \cite{weiss2018}.
	The result can straightforwardly be extended to INAR models with a general order $p$.

\subsection{Double-bounded auto-regressive model}
\label{SEC: beta}

As a second illustration, we consider an application of the asymptotic results to double-bounded time series processes. We study a process that takes values in the unit interval $[0,1]$, however, we note that the same results apply to the generic bounds $[L,U]$ as the observable process can be transformed to lie in the unit interval. We consider  the following specification for the  conditional mean and pseudo-variance of the PVQMLE
\begin{align} \label{beta autoregression}
	\lambda_t &= \omega_1 + \alpha_1 Y_{t-1} + \beta_1 \lambda_{t-1} ,\\
	\nu^*_t  & =\frac{\mu_{t}(1-\mu_{t})}{1+\phi}, \qquad \mu_t= \omega_2 + \alpha_2 Y_{t-1} + \beta_2 \mu_{t-1}, \nonumber
\end{align} 
where the double-bounded nature of the data requires $0 < \omega_i + \alpha_i + \beta_i <1$ for $i=1,2$ and $\phi >0$. If the observable variable $Y_t$ follows a conditional beta distribution with mean $\lambda_t$ and dispersion parameter $\phi$, $Beta(\lambda_t, \phi)$, then the conditional variance will take the form defined in \eqref{beta autoregression} with $\mu_t=\lambda_t$.  We assume this beta process as data generating process. 
\begin{theorem} \label{Thm. stationarity beta}
	Assume the process $\left\lbrace Y_t, \lambda_t \right\rbrace_{t \in \Z} $ is generated by $Y_t|\Fb_{t-1} \sim Beta(\lambda_t, \phi)$ with conditional mean specified as in \eqref{beta autoregression} and $\omega_1 + \alpha_1 + \beta_1 <1$. Then, the process admits
	a strictly stationary and ergodic solution with finite moments of any order $\E(Y_t^r)< \infty$ for all $r \geq 1$.
\end{theorem}
The results follows immediately by \citet[Thm.~2.1]{gorgi2021beta} and all moments exist since the time series is bounded. Next, we  derive the strong consistency
and asymptotic normality of the PVQML estimators of double-bounded autoregressions. We consider the case where the observations are generated from $Beta(\lambda_t, \phi)$ with mean specified as in \eqref{beta autoregression}. We study the asymptotic properties of the unrestricted PVQMLE and the restricted PVQMLE with restrictions $\omega_2 =\omega_1$, $\alpha_2=\alpha_1$ and $\beta_2=\beta_1$. 
	\begin{theorem} \label{Thm. can beta}
	Let $\{Y_1,\dots,Y_T\}$ be generated by a beta autoregressive process with conditional distribution $Y_t|\Fb_{t-1} \sim Beta(\lambda_t, \phi)$ where $\lambda_t$ follows the recursive equation in \eqref{beta autoregression}. 
	Consider PVQMLEs for the parameter vector $\theta=(\omega_{1},\alpha_{1},\beta_{1}, \omega_{2},\alpha_{2},\beta_{2})'$ based on the pseudo-variance $\nu^*_t$ specified   in \eqref{beta autoregression}. 
Furthermore, assume that $\theta_0 \in \Theta$, where $\Theta$ is a compact parameter set such that   $\omega_i >0$, $\alpha_i >0$, $\beta_i \geq 0$, $\omega_i + \alpha_i + \beta_i <1$, $\phi >0$, for $i=1,2$ and  for any $\theta \in \Theta$. Then, the unrestricted PVQMLE \eqref{unconstrained}
		and the restricted PVQMLE \eqref{constrained}  with restrictions $\omega_2 =\omega_1$, $\alpha_2=\alpha_1$ and $\beta_2=\beta_1$ are strongly consistent. 
		Assume further that $\theta_0 \in \dot{\Theta}$. Then, both PVQMLEs are asymptotically normally distributed with asymptotic covariance matrix given in Theorems~\ref{Thm. can}-\ref{Thm. constrained can}. 
	\end{theorem}

\section{Efficiency of the PVQMLE}
\label{SEC: Efficiency}

\subsection{Comparison to alternative estimators}
\label{SUBSEC: Model Uncons}
In this section, we show that the unrestricted PVQMLE  achieves the same asymptotic variance of existing estimators. 
Consider the unrestricted PVQMLE depicted in Theorem~\ref{Thm. can}. The partition of the score related
to the mean parameter $\psi$ is
\begin{equation}
	\tilde s_t^{(\psi)}(\theta)= \frac{Y_t- \tilde \lambda_t(\psi)}{\tilde \nu^*_t(\gamma)} \frac{\partial \tilde \lambda_t(\psi)}{\partial \psi}.
	\label{sqmle mean}
\end{equation}
We compare \eqref{sqmle mean} with some alternative semi-parametric estimators presented in the literature.

Consider the two-stage Weighted Least Squares (WLSE) of \cite{francq_2021two} defined as
\begin{equation*}
	\hat \psi_W = \argmax_{\psi \in \Psi } \frac{1}{T} \sum_{t=1}^{T} \tilde ls_t  (\psi,  \hat w_t )\,, \quad   
	\tilde  ls_t(\psi, \hat w_t ) = -\frac{[Y_t-\tilde \lambda_t(\psi)]^2}{ \hat w_t},
\end{equation*}
where $\hat w_t$ is a first-step estimator of the set of weights $w_t$.
The resulting score of the WLSE is
\begin{equation}
	\tilde s_t(\psi, \hat w_t )= \frac{Y_t- \tilde \lambda_t(\psi)}{\hat w_t} \frac{\partial \tilde \lambda_t(\psi)}{\partial \psi}.
	\label{wlse}
\end{equation}
Since it is well-known that the conditional variance is the optimal weight for the WLSE, the same authors set $ w_t = \nu^*_t (\xi) = \nu^*  (Y_{t-1}, Y_{t-2}, \dots ; \xi)$  by defining a functional form for a pseudo-variance, 
where the parameters $\xi$ may also contain $\psi_0$ or parts of it. The corresponding first-step estimated weights are 
$\hat w_t = \tilde \nu^*_t (\hat \xi)$, where $\hat \xi$ represents the first-step estimate of the parameter $\xi$.

	Another related estimator is the Estimating Function (EF) approach for dynamic models that has been recently introduced by \cite{francq2023optimal}. In the case where only the conditional mean is correctly specified, the EF equation can be written as a slightly modified version of \eqref{wlse}, by setting $\hat w_t = \tilde \nu^*_t (\tilde \xi)$ and $\tilde \xi =(\psi, \hat \zeta)$ where $\hat \zeta$ are first-step estimates of parameters not in common with the mean equation. The estimator is then defined as the solution of the following system of equation $ \sum_{t=1}^{T} \tilde s_t(\psi, \hat w_t )=0$.

Consider the general QMLE  of \cite{wedderburn1974quasi} and  \cite{gourieroux1984pseudo} based on the exponential family of quasi-likelihoods defined as
\begin{equation*}
	\hat \psi_Q = \argmax_{\psi \in \Psi } \tilde l_T(\psi)\,,  
\end{equation*}
where the log-quasi-likelihood $ \tilde l_T(\psi)$ is a member of  the one-parameter exponential family with respect to $\tilde \lambda_t(\psi)$. The corresponding score is given by 
\begin{equation}
	\tilde s_t(\psi)= \frac{Y_t- \tilde \lambda_t(\psi)}{\tilde \nu_t(\psi)} \frac{\partial \tilde \lambda_t(\psi)}{\partial \psi} \,,
	\label{qmle}
\end{equation}
where the conditional variance $\tilde \nu_t(\psi)$ is typically a function of the mean, i.e.~$\tilde\nu_t(\psi)=h( \tilde\lambda_t(\psi))$ for some function $h(\cdot)$.  For example,  selecting the Poisson quasi-likelihood yields   $\tilde \nu_t(\psi) = \tilde \lambda_t(\psi)$  \cite[]{fra2016}, see \citet[Sec.~2.2]{francq_2021two} for other examples.  

The expressions of the scores in \eqref{sqmle mean}-\eqref{qmle} highlight how the unrestricted PVQMLE is related to WLSE, EF estimator and the QMLE based on the exponential family. The main difference between the unrestricted PVQMLE and the QMLE with score in \eqref{qmle} is that the QMLE only considers the specification of the conditional mean and the conditional variance  is a function of the  conditional mean that is implied by the selected distribution in the exponential family. On the other hand, the unrestricted PVQMLE differs from the WLSE  as the parameters are estimated jointly instead of a multi-step estimation. A similar difference applies between the unrestricted PVQMLE and the EF approach, which also estimates some of the variance parameters in a first stage.   The unrestricted PVQMLE, the QMLE, the WLSE and the EF estimator enjoy the same consistency property   for the mean parameters $\psi_0$ irrespective of the correct specification of the  conditional variance. Furthermore, when they have the same specification of the conditional pseudo-variance, these estimators are asymptotically equivalent.



	
	
	
	\begin{corollary} \label{Cor. equivalence}
		Assume Theorem~\ref{Thm. can} holds. Moreover, 
		suppose the WLSE \eqref{wlse} with $w_t=\nu^*_t(\gamma^*)$ is consistent and asymptotically normal with limiting variance $\Sigma_W$.
		Then the unrestricted PVQMLE in \eqref{unconstrained}  is asymptotically as efficient as the WLSE, meaning that $\Sigma_\psi = \Sigma_W$.
		In addition, if $\nu^*_t(\cdot)=\nu_t(\cdot)$, then $\Sigma_W = \Sigma_\psi = I_\psi^{-1} $.
	\end{corollary}
	
	The result in  Corollary \ref{Cor. equivalence}   follows  immediately from Theorem~\ref{Thm. can} and Corollary~\ref{Cor. correct variance}. 
	Since the EF estimator still involves a two-step procedure, it is not surprising to see that   the EF estimator has  
		the same efficiency as the WLSE \citep{francq2023optimal}. Therefore, the results of Corollary \ref{Cor. equivalence} also hold for  the EF approach.
	 We also note that if Corollary~\ref{Cor. equivalence} holds then also Corollaries  2.1-2.3 in \cite{francq_2021two} hold for the unrestricted PVQMLE. This has two direct consequences:  (i) if the variance is well-specified, the unrestricted PVQMLE is asymptotically more efficient than the QMLE of $\psi_0$, if the variance implied by the exponential family is not the true one, and 
	(ii) if the conditional distribution of $Y_t$ comes from the exponential family, then the well-specified PVQMLE is asymptotically as efficient as the MLE of  $\psi_0$.
	
	We note that the comparison discussed so far only concerns the unrestricted PVQMLE. This asymptotic equivalence of the PVQMLE with respect to the WLSE, the EF estimator and the QMLE does not hold for the restricted PVQMLE. This can  be noted from the form of the score function given in equation \eqref{score} and the fact that the partial derivative of $\tilde \nu^*_t(\gamma)$ with respect to $\psi$ is no longer equal to zero.  This  partial derivative is non-zero also in the EF approach for the special case of correctly specified conditional variance. However, even in this special case of correct specification of the conditional variance, PVQMLEs differ from the  EF approach as  the latter assumes that $\tilde \nu_t^*$ only depends on the parameter $\psi$, i.e.~no additional  free parameters are allowed in the conditional variance equation, which  is instead included in our approach. Below we discuss how the  restricted PVQMLE can achieve higher efficiency compared to the unrestricted PVQMLE.

	
	%
	
	\subsection{Some results on the efficiency of PVQMLE}
	\label{SUBSEC: Improved efficiency}
	
	Given that the  PVQMLE with distinct parameters on mean and  pseudo-variance is asymptotically equivalent to the WLSE for the mean parameters $\psi_0$ (Corollary~\ref{Cor. equivalence}), it may be expected that if the mean and pseudo-variance equations share common parameters in $\theta$, i.e. $\psi_0$ and $\gamma^*$ are not completely distinct so that $\theta_0 \in \Theta_R$, then the restricted PVQMLE in  \eqref{constrained}  could show improved efficiency  over the unrestricted PVQMLE and the WLSE. It is not straightforward to prove this result in general but for the following special cases it is verified.
	
	\begin{enumerate}[label=\textbf{A13}]
		\item $E(Y_t^4|\mathcal{F}_{t-1}) <\infty$ almost surely. \label{Ass. conditional moments}
	\end{enumerate}
	
	\begin{enumerate}[label=\textbf{A14}] 
		
		\item \label{Ass. efficiency}  Set $p=k=1$ and $Y_t| \Fb_{t-1} \sim q(\lambda_t, \nu_t)$ where $q(\cdot)$ has kurtosis $\leq 3$. One of the following conditions holds:
		\begin{enumerate}[label=\textbf{A14.\alph*}]
			\item $q(\cdot)$ is symmetric. 
			\item The first derivatives of the functions $\lambda_t(\psi_0)$ and $\nu_t(\gamma_0)$ have the (opposite) same sign and $q(\cdot)$ is  (positive) negative skewed. 
		\end{enumerate}

	\end{enumerate}

	\begin{proposition} \label{Thm. normal eff}
		Assume that Assumptions \ref{Ass. stationarity}-\ref{Ass. efficiency} hold with $\nu^*_t(\gamma^*)=\nu_t$. 
		Moreover, 
		suppose that the WLSE in \eqref{wlse} with $w_t=\nu_t$ is consistent and asymptotically normal with asymptotic variance $I^{-1}_\psi$.
		Then, the  restricted PVQMLE in \eqref{constrained}  is asymptotically more efficient than the unrestricted PVQMLE and the WLSE, i.e.~$I^{-1}_\psi - \Sigma_{R}$ is positive semi-definite.
	\end{proposition}
	
	The conditions stated in Assumption~\ref{Ass. efficiency} can be  somewhat restrictive, however, we note that they  are  only sufficient conditions.  In general, it is not straightforward to derive sharper theoretical conditions under which the restricted  PVQMLE  is more efficient than the unrestricted PVQMLE. However, for specific models, we  can appeal to numerical methods to obtain the asymptotic covariance matrix of the two estimators and  evaluate their  relative efficiency. 

	We  consider the  INAR($1$) model  in \eqref{inar} with binomial thinning and Poisson error distribution as an example. 	The unrestricted  PVQMLE $\hat \psi$ is based on the following conditional mean and pseudo-variance equations 
	\begin{equation}
		\lambda_t(\psi)=a  Y_{t-1} + \omega_1\,, \quad \quad  \nu_t^*(\gamma)=b Y_{t-1} + \omega_2\,,
		\label{second order linear}
	\end{equation}
	where $\psi'=(a,\omega_1)$ and $\gamma'=(b,\omega_2)$.
	Instead, the restricted PVQMLE $\hat \psi_R$ imposes the restrictions $b=a(1-a)$ and $\omega_2=\omega_1 = \omega$.

\begin{figure}[h!]
		\centering
		\subfloat
		{{\includegraphics[width=9cm]{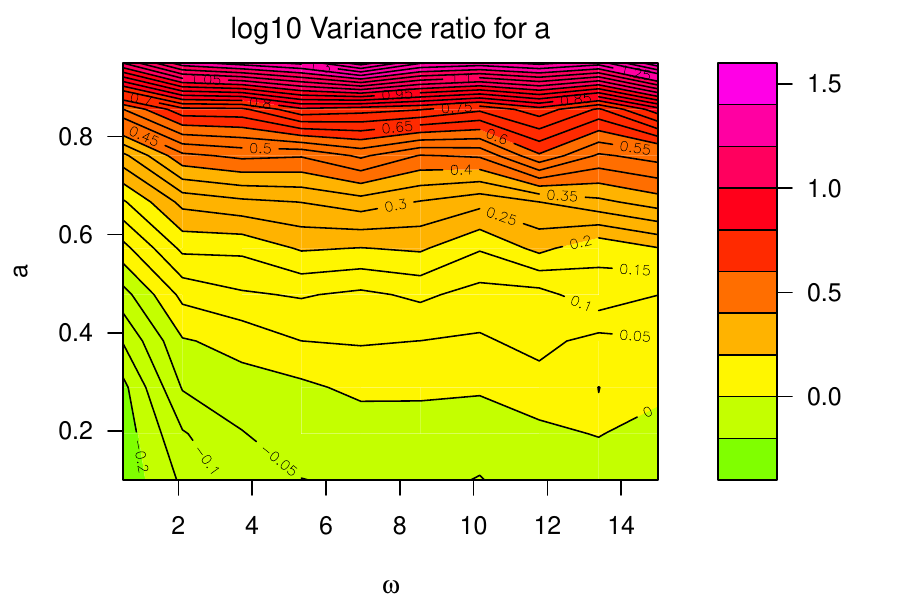} }}%
		\subfloat
		{{\includegraphics[width=9cm]{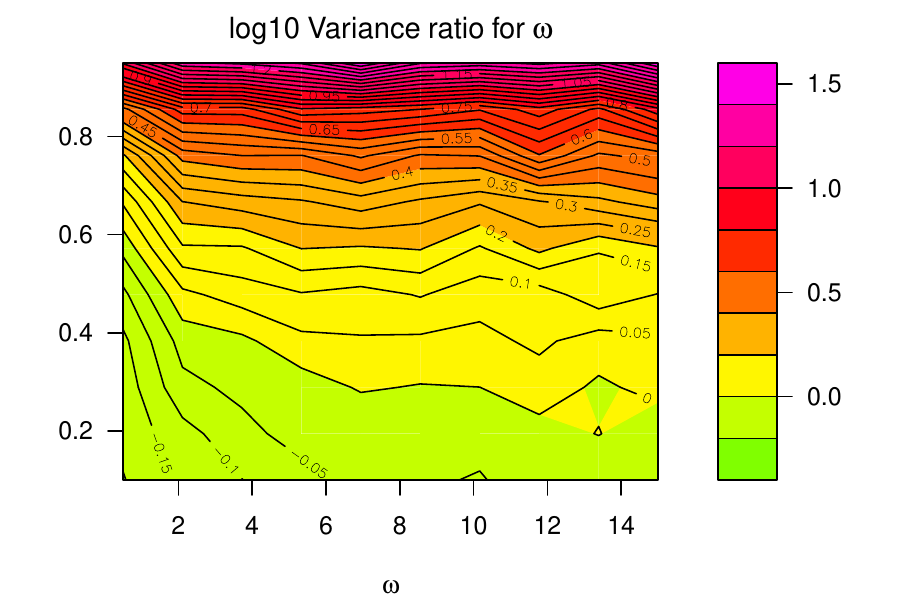} }}%
		\caption{Contour plots of $\log_{10}$-variance ratios for the INAR coefficients. Left: ratio $\log_{10}[Var(\hat a)/Var(\hat a_R)]$ plotted for several values of $a$ and $\omega$. Right: ratio $\log_{10}[Var(\hat \omega)/Var(\hat \omega_R)]$ plotted for several values of $a$ and $\omega$. The green area indicates a variance ratio smaller than one. }
		\label{Fig: theoretical efficiency INAR 3d}
	\end{figure}

\begin{figure}[h!]
		\centering
		\includegraphics[width=0.7\linewidth]{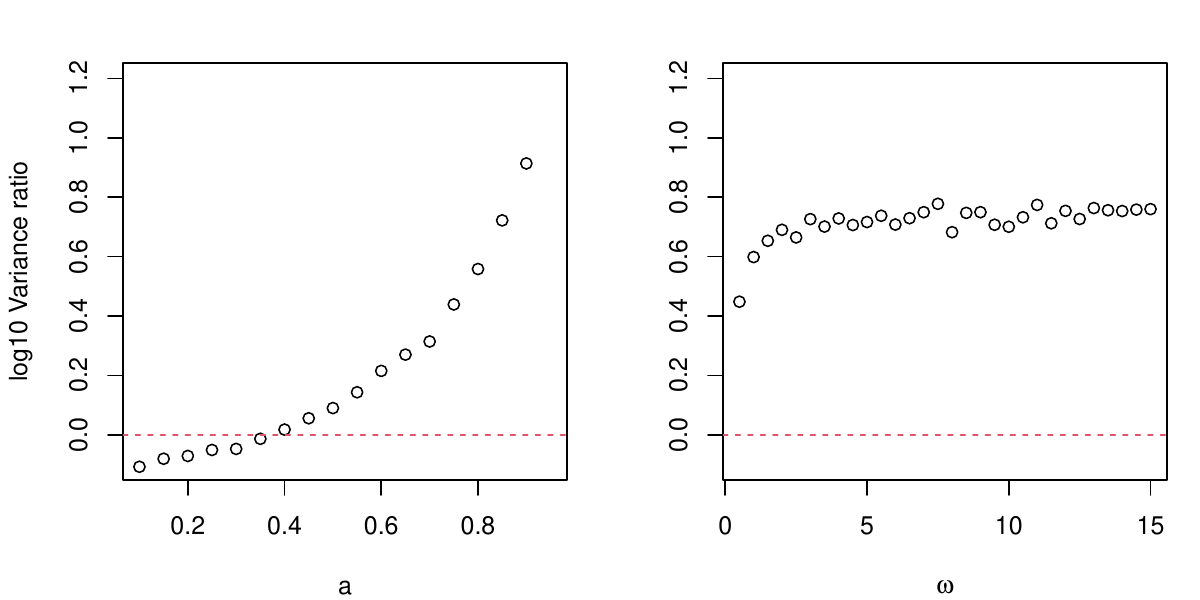}
		\caption{$\log_{10}$-variance ratios plots for the INAR coefficients. Dashed red line: y-axis=0. Left: ratio $\log_{10}[Var(\hat a)/Var(\hat a_R)]$ plotted for several values of $a$ and $\omega=3$. Right: ratio $\log_{10}[Var(\hat \omega)/Var(\hat \omega_R)]$ plotted for several values of $\omega$ and $a=0.85$.}
		\label{Fig: theoretical efficiency INAR}
	\end{figure}

	We focus on the analysis of the asymptotic variances of these estimators. To this aim, we simulate a long time series ($T=10,000$) from the  INAR($1$) process (binomial thinning and Poisson errors) for different values of the parameters $a$ and $\omega_1$ over a grid. The asymptotic covariance matrices of the two estimators are computed by approximating their expectations with the corresponding sample means.  	Figure \ref{Fig: theoretical efficiency INAR 3d} reports an heatmap plot of the ratio (in $\log_{10}$ scale)  between the asymptotic variance of the unrestricted and the restricted PVQMLEs for the parameter estimates of $a$ and $\omega_1$. The regions  of the parameter set  where the $\log_{10}$-variance ratio is  greater than zero, i.e. variance ratio is greater than one, indicate the parameter values for which the restricted estimator is more efficient of the unrestricted one, and vice versa.   The pictures suggest that the restricted estimator  $\hat \psi_R$ is more efficient than the unrestricted estimator $\hat \psi $ in most cases, except when $a$ and $\omega_1$ are close to zero. Furthermore, the lack of efficiency of the restricted PVQMLE  in the green areas is showed to be minimal. For example, a $\log_{10}$-variance ratio around $-0.05$ indicates a variance ratio around $0.9$. Therefore, for small values of $a$ and $\omega_1$ the two estimators are essentially equivalent. Instead, for larger values of $a$ and $\omega_1$,   the variance ratio gets substantially larger with the unrestricted PVQMLE estimator having up to 30 times larger variance of the restricted one.  This is further illustrated in Figure \ref{Fig: theoretical efficiency INAR}, which displays  a graph of cross-section of the  $\log_{10}$-variance ratio for some fixed values of $a$ and $\omega_1$.

	Another way to grasp the intuition behind the improved efficiency of the restricted PVQMLE   comes from the literature on saddlepoint approximations   \cite[]{daniels1954saddlepoint}. Saddlepoint approximations are used to approximate a density function with a  function that is based on the cumulant generating function of the data, which is typically called saddlepoint density.   \cite{pedeli2015likelihood} show that the conditional saddlepoint density can approximate the conditional density of the INAR($p$) model in \eqref{inar} to a certain degree of accuracy.
	It is not hard to see that the conditional saddlepoint density is approximately equal to the pseudo-variance quasi-likelihood  in \eqref{quasi likelihood} with correctly specified  variance \cite[Sec.~3.4]{pedeli2015likelihood}.
	Therefore, when the variance is correctly specified, the restricted   PVQMLE of the INAR($p$) model  is close to the maximizer of the log-likelihood obtained by the saddlepoint density, which in turn is expected to get closer to the MLE   as $\lambda_t \to \infty$.   This is confirmed empirically from the results in Figures \ref{Fig: theoretical efficiency INAR 3d} and \ref{Fig: theoretical efficiency INAR}, where the efficiency of restricted PVQMLE  over the unrestricted PVQMLE    grows as $a,w \to \infty$ i.e.~where restricted PVQMLE    approximates more accurately the MLE. We conjecture that similar results may apply also to other models.
	For the case of independent observations, \cite{goodman2022asymptotic} has recently shown that  the approximation error in using saddlepoint approximation is negligible compared to the inferential uncertainty  inherent in the MLE. Although the literature is still under development, these arguments provide reliable evidence on the higher asymptotic performance of  restricted PVQMLEs compared to the unrestricted one and other quasi-likelihood methods presented in Section~\ref{SUBSEC: Model Uncons}.

	Finally, we consider a simulation study to assess the small sample properties of PVQMLEs in comparison with several other alternative estimators.  The study consists of 1,000 Monte Carlo replications where we generate data from the Poisson INAR($1$) process and estimate the mean parameter vector $\psi$.  We consider several PVQMLEs based on different restrictions of the variance parameter vector $\gamma$. The unrestricted PVQMLE $\hat \psi$ is based on the mean and pseudo-variance equations in \eqref{second order linear}. The first restricted PVQMLE  $\hat \psi_{R_1}$ imposes the restriction $R_1:  b=a(1-a)$, the second restricted PVQMLE  $\hat \psi_{R_2}$ imposes the restriction $R_2:  \omega_2=\omega_1$, and the third  restricted PVQMLE  $\hat \psi_{R_3}$ imposes the restriction $R_3:  b=a(1-a), \; \omega_2=\omega_1$. Furthermore, we consider the QMLE based on the Poisson quasi-likelihood $\hat \psi_Q$, the conditional least squares estimator  (CLSE) $\hat \psi_{LS}$,  the WLSE in \eqref{wlse} with weights $\hat{w}_t=\hat{a}_{LS} (1 - \hat a_{LS} )Y_{t-1} + \hat \omega_{LS}$ where $(\hat \omega_{LS}, \hat a_{LS})^\prime = \hat \psi_{LS}$ are first-step estimates obtained from the CLSE, and  the  unfeasible WLSE $ \hat \psi_{WUN} $  with weights given by the true conditional variance. 
	Finally, we  also include   the  MLE $\hat \psi_{ML}$ for comparison purposes.
	 The results of the simulation study are reported in Table \ref{Tab. INAR}.

\begin{table}[h!]
	\centering
	\caption{Bias and RMSE of estimators of the mean parameters when the data generating process is an INAR($1$) with $a=0.85$ and $\omega=3$, and sample size $T = \{100,500,2000\}$.}
	\scalebox{0.75}{
		\begin{tabular}{cccccccccccccccccc}
			\noalign{\smallskip} \hline \noalign{\smallskip}
			& \multicolumn{6}{l}{$T = 100$}& \multicolumn{6}{l}{$T = 500$} &  \multicolumn{5}{l}{$T = 2000$} \\
			\cline{2-6}
			\cline{8-12} 
			\cline{14-18} 
			& \multicolumn{3}{l}{$\omega_1$} & \multicolumn{3}{l}{$a$} &
			\multicolumn{3}{l}{$\omega_1$} & \multicolumn{3}{l}{$a$} & \multicolumn{3}{l}{$\omega_1$} & \multicolumn{2}{l}{$a$} \\ 
			\cline{2-3}
			\cline{5-6}
			\cline{8-9}
			\cline{11-12}
			\cline{14-15}
			\cline{17-18}
			Est. & Bias & RMSE & & Bias & RMSE & & Bias & RMSE & & Bias & RMSE & & Bias & RMSE & & Bias & RMSE  \\
			\noalign{\smallskip}\hline \noalign{\smallskip}
			$\hat \psi_Q$ & 0.7462 & 1.4348 && -0.0389 & 0.0736 && 0.1496 & 0.5218 && -0.0077 & 0.0262 && 0.0323 & 0.2425 && -0.0016 & 0.0121 \\ 
			$\hat \psi_{LS}$ & 0.7453 & 1.4391 && -0.0389 & 0.0739 && 0.1503 & 0.5217 && -0.0077 & 0.0262 && 0.0290 & 0.2401 && -0.0014 & 0.0119 \\ 
			$\hat \psi_{W}$ &  0.7382 & 1.4319 && -0.0385 & 0.0735 && 0.1475 & 0.5175 && -0.0076 & 0.0260 && 0.0301 & 0.2392 && -0.0015 & 0.0119 \\ 
			$\hat \psi_{WUN}$ &  0.7377 & 1.4318 && -0.0385 & 0.0735 && 0.1474 & 0.5173 && -0.0076 & 0.0260 && 0.0300 & 0.2393 && -0.0015 & 0.0119 \\ 
			$\hat \psi$ &  0.7203 & 1.4235 && -0.0376 & 0.0731 && 0.1452 & 0.5166 && -0.0075 & 0.0260 && 0.0295 & 0.2388 && -0.0015 & 0.0119 \\ 
			$\hat \psi_{R_1}$ &  0.7050 & 1.3837 && -0.0368 & 0.0710 && 0.1417 & 0.4985 && -0.0073 & 0.0250 && 0.0305 & 0.2314 && -0.0015 & 0.0115 \\ 
			$\hat \psi_{R_2}$ &  0.5913 & 1.1980 && -0.0313 & 0.0620 && 0.1316 & 0.5051 && -0.0068 & 0.0255 && 0.0246 & 0.2332 && -0.0012 & 0.0115 \\ 
			$\hat \psi_{R_3}$ & 0.0311 & 0.4586 && -0.0028 & 0.0235 && -0.0010 & 0.2036 && -0.0002 & 0.0101 && 0.0027 & 0.1011 && -0.0001 & 0.0049 \\ 
			$\hat \psi_{ML}$ &  0.0317 & 0.4551 && -0.0028 & 0.0234 && -0.0002 & 0.2018 && -0.0002 & 0.0100 && 0.0029 & 0.1009 && -0.0001 & 0.0049 \\ 
			\hline
		\end{tabular}
	}
	\label{Tab. INAR}
\end{table}
	
	Since the PVQMLE without constraints on the first two moments is asymptotically equivalent to the WLSE, it can be expected that the restricted PVQMLE where suitable constraints corresponding to the true model are imposed should show improved performances  over the other quasi-likelihood-type estimators. Indeed, from Table~\ref{Tab. INAR} it can be seen that QMLE, CLSE, WLSE and unrestricted PVQMLE   of model \eqref{second order linear} share similar performances both in terms of bias and RMSE. Instead, a partial specification of the true constraints underlying the model in $\hat \psi_{R_1}$ and $\hat \psi_{R_2}$ already leads to an improvement with respect to the other estimation techniques; such improvement becomes substantial in $\hat \psi_{R_3}$ where all the correct constraints are considered. 
	Moreover, this last restricted PVQMLE has comparable performance to the MLE. This is important since when $p\gg 1$ the MLE can become hard to compute and therefore our approach is a valid alternative.

	\section{Testing restrictions}
	\label{SEC: Specification tests}
	
	In the Section~\ref{SEC: Asymp. theory}, we have seen that correctly identified constraints on mean and pseudo-variance equations can deliver a restricted PVQMLE with improved efficiency. 
	In this section, we develop a test based on the   unrestricted estimator in \eqref{unconstrained} which allows us to test the validity of the restriction $S\gamma=g(\psi)$. We define $r(\theta)=S\gamma-g(\psi)$ and we denote with $\Sigma(\theta_0)=H^{-1}(\theta_0) I(\theta_0) H^{-1}(\theta_0)$  the asymptotic  covariance matrix of the entire unrestricted  estimator vector $\hat{\theta}$.  Moreover, consider the following plug-in estimators of   $H(\theta_0)$ and $I(\theta_0)$ given by  $\tilde{H}_T(\hat \theta)=T^{-1}\sum_{t=1}^{T} -\partial^2 \tilde l_t(\hat \theta)/\partial \theta \partial \theta^\prime$ and $\tilde{I}_T(\hat \theta) = T^{-1}\sum_{t=1}^{T} \tilde s_t(\hat \theta)  \tilde s^\prime_t(\hat \theta)$, respectively.
	The following result holds.
	
	\begin{proposition} \label{Thm. test}
		Assume that the assumptions of Theorem~\ref{Thm. can} hold. Consider the test $H_0: r(\theta_0) = 0$ versus $H_1: r(\theta_0) \neq 0$ where the function $r(\cdot)$ is continuously differentiable. Let $R(\theta) = \partial r(\theta)/\partial \theta^\prime$.   Then, under $H_0$, as $T\to\infty$
		\begin{equation*}
			W_T = T r^\prime(\hat \theta) \big[ R (\hat \theta) 
			\Sigma(\theta_0)
			R ^\prime (\hat \theta) \big]^{-1} r(\hat \theta) \xrightarrow{d} \chi^2_r,
		\end{equation*}
		where we can estimate $\Sigma(\theta_0)$ by $\tilde{\Sigma}_T(\hat \theta)= \tilde{H}_T^{-1}(\hat \theta) \tilde{I}_T(\hat \theta)  \tilde{H}_T^{-1}(\hat \theta)$.
	\end{proposition}
	
	The result follows immediately by the multivariate delta method,  the continuous mapping theorem and standard asymptotic convergence arguments.
	Proposition~\ref{Thm. test} provides us a  testing procedure for $H_0: \theta_0 \in \Theta_R$ versus $H_1: \theta_0 \notin \Theta_R$.  It is worth nothing that the hypothesis test depicted in Proposition~\ref{Thm. test} does not require the variance of the model to be correctly specified. In the special case in which the  pseudo-variance is correctly specified, then the   test can be interpreted as  a test of correct specification.

	For example, consider the INAR($1$) model in \eqref{inar}  with conditional mean and pseudo-variance equations as defined in equations \eqref{inar mean}-\eqref{inar pseudo-var}.
	We may consider the following test
	\begin{equation} \label{test binomial thin}
		H_0: b = a (1-a) \quad  \text{vs} \quad H_1: b \neq  a (1-a)\,,
	\end{equation}
	which is a test for the assumption of a binomial thinning operator $`\circ$'.  This follows from the definition of the INAR model in \eqref{inar} as the autoregressive coefficient of the variance takes the form $b=a(1-a)$ under the assumption of binomial thinning. Alternative thinning specifications can be tested leading to a different form of the autoregressive variance parameter $b$, see \cite{latour1998existence} for the properties of INAR models with a general thinning specification. For instance, if we have a Poisson distribution for the thinning   operator we have the restriction $b=a$.
	The corresponding test is 
	\begin{equation} \label{test pois thin}
		H_0: b = a  \quad  \text{vs} \quad H_1: b \neq  a,
	\end{equation}
	which assesses the validity of the assumption of equidispersion in the thinning operator versus either overdispersion or underdispersion.

	\begin{table}[h!]
		\centering
		\caption{Empirical size for test in  \eqref{test pois thin}.
			The model considered under $H_0$ is an INAR($1$) model with Poisson thinning as well as Poisson error with parameter values $a=0.75$ and  $\omega=1$.}
		\scalebox{0.95}{
			\begin{tabular}{cccccc}
				\hline
				&	\multicolumn{4}{c}{$T$} \\
				\cline{2-6}
				Nominal size & 100 & 250 & 500 & 1000 & 2000 \\ 
				\noalign{\smallskip}\hline \noalign{\smallskip}
0.1000	& 0.1222 & 0.1222 & 0.1140 & 0.1060 & 0.0986 \\ 
 	0.0500 & 0.0720	& 0.0642 & 0.0582 & 0.0514 & 0.0518 \\ 
	0.0100 & 0.0202	& 0.0142 & 0.0114 & 0.0122 & 0.0128 \\ 
				\hline
			\end{tabular}
		}
		\label{Tab: empirical size pois vs nb}
	\end{table}

	\begin{figure}[h!]
		\centering
		\begin{tabular}{c}
			\includegraphics[width=0.8\linewidth]{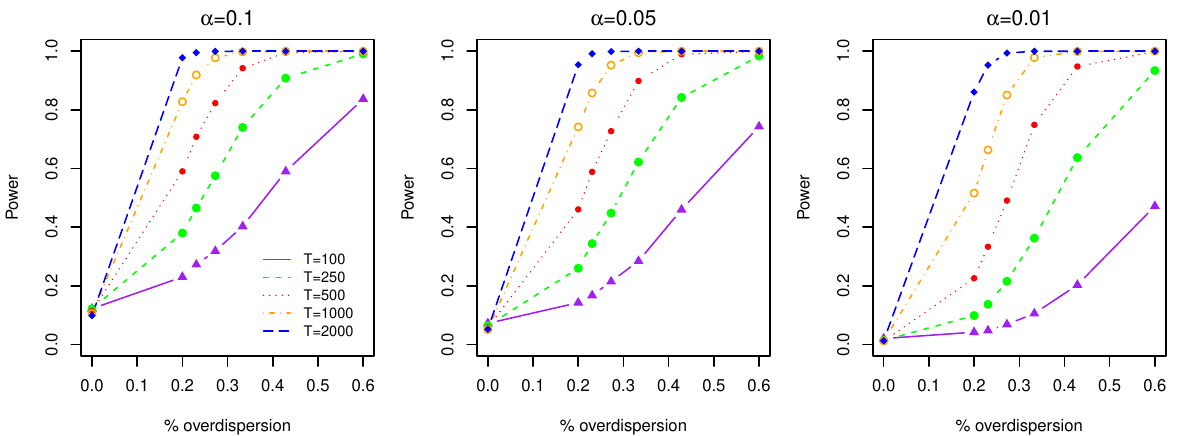}
		\end{tabular}	
		\caption{Empirical power for test in \eqref{test pois thin}. The true parameter values of the  INAR($1$) model with negative binomial thinning and Poisson error are $a=0.75$ and $\omega = 1$. The value of the dispersion parameter $v$ changes  as indicated in the horizontal axis through the \% of overdispersion: $1-a/(a+a^2/v)$.}
		\label{Fig: power pois vs nb}
	\end{figure}	
	
	We carry out  a simulation study with $5000$  Monte Carlo replications to assess the empirical size and power of the test of the parameter restrictions for the INAR($1$) model. We consider the hypothesis in \eqref{test pois thin}. To assess the size of the test we simulate under $H_0$ from a model with Poisson thinning operator and a Poisson distribution of the error term. Table~\ref{Tab: empirical size pois vs nb} reports the results on the empirical size of the test. We can see that the test is slightly oversized for the smallest sample size, though still close to the nominal level, and it quickly becomes correctly sized as the sample size increases. Next, we evaluate the power of the test by simulating under the alternative. We consider a  negative binomial   thinning  specification such that $a \circ N$ has a negative binomial distribution with mean $a N$  and variance $ bN$, $b = a+a^2/v$, where $v$ is the dispersion parameter of the negative binomial. We note that this generates overdipersion in the thinning as $b = a+a^2/v>a$ and the smaller the parameter $v$ the more the overdispersion.  Figure~\ref{Fig: power pois vs nb} shows the power of the test in \eqref{test pois thin} to reject the null hypothesis. As expected, we  see that the power increases  as the relative  overdispersion $1-a/b$ increases ($v$ decreases) and as the sample size increases.  Overall, the results show how the test has appropriate size and it has power against alternative hypotheses.

We also report additional simulation results in Appendix~\ref{Appendix B further numerical results} that consider alternative data generating processes.
First, we evaluate the robustness of the described test statistic by repeating the same simulation study with the inclusion of an outlier defined as 3 times the standard deviation of the observations plus their sample mean. The results show that the test is slightly oversized but the empirical size is still in line with nominal values. 
 Moreover, the power of the test converges to  1 at a slightly slower rate with the increasing sample size but it still performs satisfactorily.
Second, we evaluate the test in case of near-unit root. The test seems conservative in this case, which can be due to the finite sample  distribution of the estimators being  different from the normal distribution near the unit-root boundary. The results on power indicate an adequate rejection rate when the sample size is large enough.
Third, we  evaluate the power of the equidispersion test under a different thinning specification. There are several thinning specifications available in the literature, see \cite{ristic2013geometric}, \cite{ilic2016geometric}, \cite{borges2016geometric}, \cite{nastic2017geometric},  \cite{borges2017generalised}, and \cite{bourguignon2018new}, amongst others.  We consider the Binomial-Negative Binomial  (BiNB) thinning  as in \cite{bourguignon2017inar}.   The results show that the power increases  as the relative  overdispersion of the thinning  increases. 
Finally, we consider the case of testing equidispersion of the thinning operator in an INAR(2) model.  As expected, the results are comparable  to the INAR($1$) results with slower convergence rate towards the correct nominal size and power as the sample size grows. This is due to a more complex testing problem and larger set of parameters to be estimated.

\section{Real data applications}
\label{SEC: Applications}

In this section, we present  two empirical applications where we employ  PVQMLEs. We consider the test described in Section~\ref{SEC: Specification tests} to select appropriate  parameter restrictions and compare different  PVQMLEs. The first application concerns a dataset of crime counts, where the INAR model is considered for the specification  of the conditional mean and the pseudo-variance. The second application concerns  the realized correlation between two financial assets that forms a double-bounded time series, where we consider a beta autoregression for the specification of the conditional mean and the pseudo-variance.

\subsection{INAR model for crime counts}

\begin{figure}[h!]
	\centering
	\includegraphics[width=0.9\linewidth]{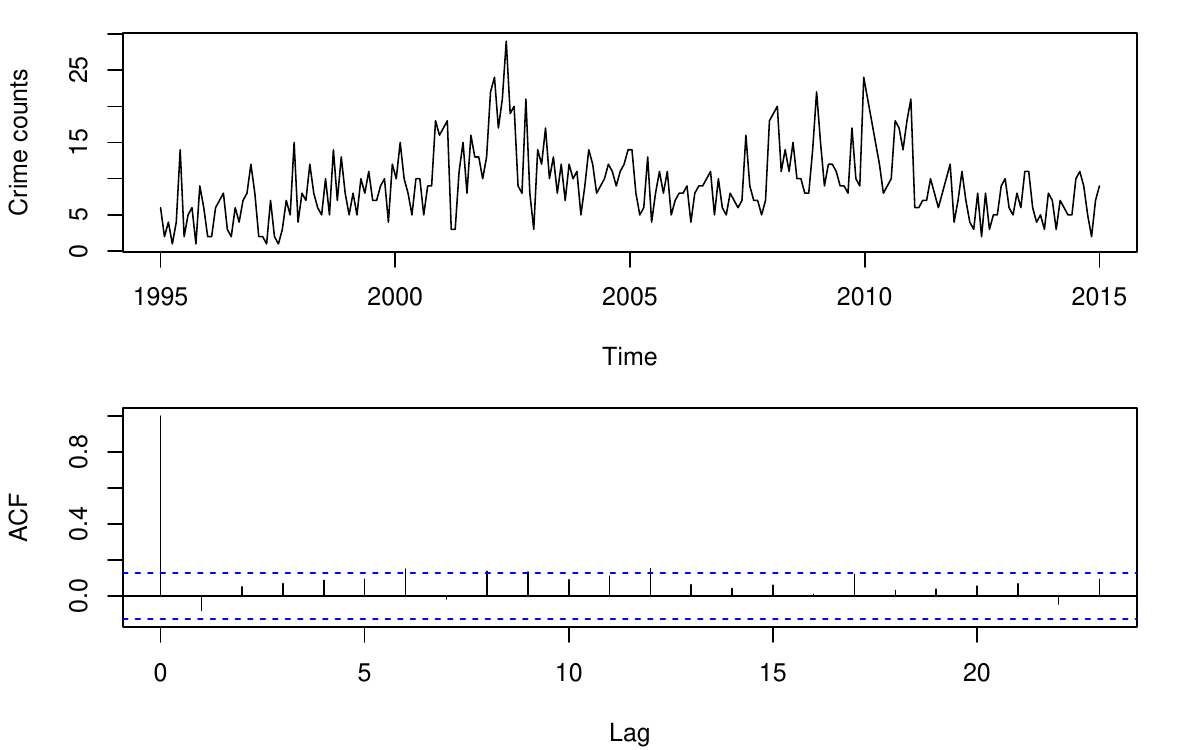}
	\caption{Monthly number of offensive conduct reports in Blacktown, Australia, from January 1995 to December 2014. The second plot represents the sample autocorrelation function of the residuals obtained from the unrestricted estimator with 95\% confidence bounds.}
	\label{Fig: crime}
\end{figure}

We consider an empirical application to the  monthly number of offensive conduct reports in the city of Blacktown, Australia, from January 1995 to December 2014. This dataset has been employed in several articles featuring the INAR($1$) model  \cite[]{gorgi2018integer, leisen2019flexible}. The time series is displayed in Figure~\ref{Fig: crime}.
In the literature, the distributional structure of the INAR innovation term $\varepsilon_t$ is typically allowed to be  flexible or left unspecified  but the thinning operator is typically considered to be binomial. We consider the test proposed in the previous section to formally test the validity of binomial thinning assumption as well as the dispersion of the error term. We obtain the unrestricted PVQMLE  for the INAR conditional mean and pseudo-variance equations in \eqref{second order linear} and test several restrictions based on the test in Proposition \ref{Thm. test}. We  test for equidispersion in the error  $H_0: \omega_1=\omega_2$,  binomial thinning $H_0: b=a(1-a)$, Poisson thinning $H_0:a=b$ and geometric thinning $H_0: b=a+a^2$. As discussed in \cite{latour1998existence}, INAR($p$) models have the same autocorrelation structure as continuous-valued AR($p$) models. In this case, we can focus on
	INAR($1$) model as the residuals obtained from the one-lag unrestricted model appear uncorrelated.

\begin{table}[h!]
	\centering
	\caption{$p$-values of the restriction tests for the  INAR($1$) model.}
	\scalebox{0.95}{
		\begin{tabular}{cccc}
			
			\hline
			$\omega_1=\omega_2$ &	 	\multicolumn{3}{c}{Thinning} \\
			\cline{2-4}
			&  binomial	 & Poisson & geometric \\ 
			\hline
			0.372 & 0.005 & 0.043 & 0.229  \\

			%
			%
			%
			%
			
			\hline
		\end{tabular}
	}
	\label{Tab: download test}
\end{table}

\begin{table}[h!]
	\centering
	\caption{PVQMLEs of the INAR($1$) model for the crime time series dataset. Standard errors in brackets.  
	}
	\scalebox{0.9}{
		\begin{tabular}{lcccc}
			\cline{2-5}
			& $\hat \omega_1$ & $\hat \omega_2$ &	$\hat a$ & $\hat b$  \\
			
			\noalign{\smallskip} \hline \noalign{\smallskip}
			
			Unrestricted & 4.559 & 6.644 & 0.509 & 1.170   \\
			& (0.520) & (2.374) & (0.058) & (0.330)  \\[0.2cm]
			binomial thinning & 6.280 & - & 0.371 & -  \\
			& (0.434) & & (0.040) &   \\[0.2cm]
			Poisson thinning & 4.820 & - & 0.524 & -  \\
			& (0.523) & - & (0.058) & -   \\[0.2cm]
			geometric thinning & 4.129 & - & 0.592 & -   \\
			& (0.500) & - & (0.059) & -   \\
			
			\noalign{\smallskip} \hline \noalign{\smallskip}
		\end{tabular}
	}
	\label{Tab: download}
\end{table}

The results of the tests are summarized in Table~\ref{Tab: download test}.  We can see that the test does not reject the hypothesis of	 equidispersion in the error $\omega_1=\omega_2$. As it concerns the tests on the thinning, the  binomial and Poisson thinning are rejected at 5\% significance level, instead, the geometric  thinning is not rejected. This indicates that there is overdispersion in the thinning and the geometric one may be appropriate to describe the degree of overdispersion.  Table~\ref{Tab: download} reports the estimation results for several PVQMLEs that are based on the different restrictions on the thinning operator. 
The standard errors are computed from the empirical counterparts of the asymptotic covariance matrices \eqref{asymp. normality} and \eqref{covariance constr} for the unrestricted and the restricted estimators, respectively.
  We can see that restricting to a binomial thinning leads to substantially biased estimates  with respect to the unrestricted PVQMLE. Instead, from the geometric thinning we do not have such bias and the estimator can be expected to have an higher efficiency.
We have also considered a BiNB thinning and it yields to equivalent estimation results as the geometric thinning. This follows as the BiNB thinning nests the geometric thinning and the estimated Bernoulli probability of the BiNB thinning is equal to zero, leading to a geometric thinning.

\subsection{Double-bounded autoregression for realized correlation}
\begin{figure}[h]
	\centering
	\includegraphics[width=0.7\linewidth]{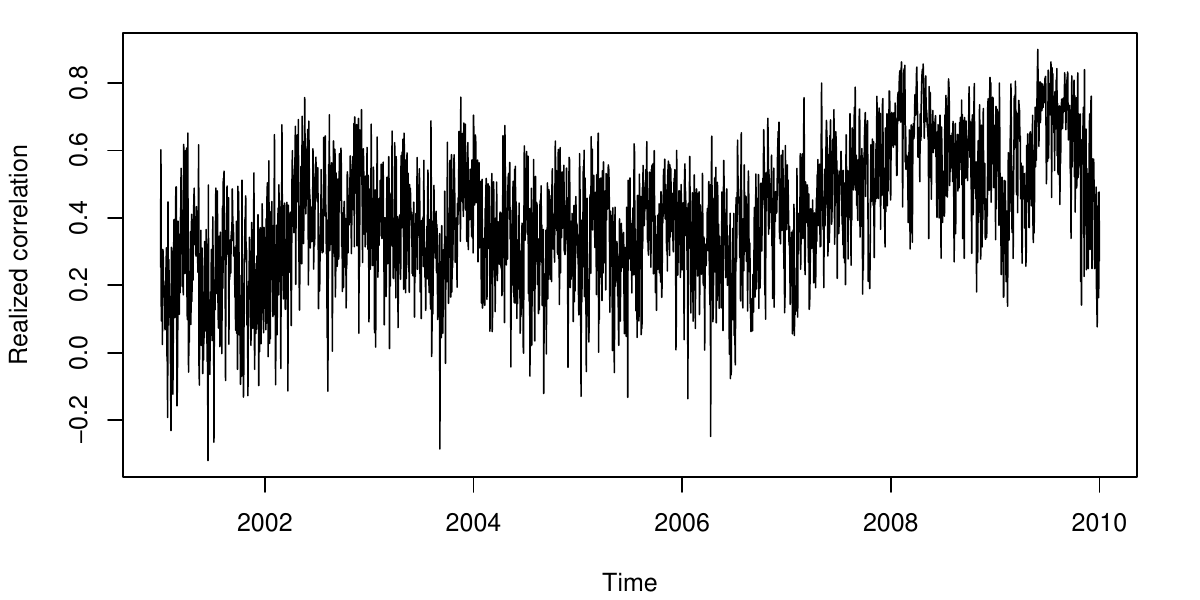}
	\caption{Daily time series of realized correlations between Boeing (BA) and Honeywell (HON) asset returns, from  January 2001 to  December 2010.}
	\label{Fig: realized corr}
\end{figure}

The second application we present concerns the modelling of daily realized correlations between  Boeing and Honeywell stocks as considered in \cite{gorgi2021beta}.    Figure \ref{Fig: realized corr} reports the plot of the time series. The sample size is $T = 2515$. Realized  correlation measures take values in the interval $[-1,1]$ and the transformation $Y_t/2+1/2$ is applied to rescale the realized correlation in the unit interval $[0,1]$. We refer to  \cite{gorgi2021beta} for a discussion on how models on the unit interval can be extended to a general interval with known bounds.

We consider the specification for the conditional mean and pseudo-variance defined in \eqref{beta autoregression}.
  Besides the unrestricted PVQMLE, we consider a restricted PVQMLE with  $\omega_2 = \omega_1$, $ \alpha_2= \alpha_1$, $ \beta_1 = \beta_2$, which implies $\mu_t=\lambda_{t}$.  These restrictions impose that the  pseudo-variance is equal to the conditional variance implied by a  beta distribution with mean parameter $\lambda_t$ and precision parameter $\phi$. In this way, we can also test the adequacy of the beta autoregression for modelling the analyzed data through the specification test  on the restriction. Table~\ref{Tab: realized corr} reports the estimation  results together with the restriction tests. 	 
We can see that the specification test rejects the null hypothesis of equality for the estimated $\alpha$  coefficients. For the same reason also the null assumption of the combined joint test is rejected. However, the null hypothesis is instead not rejected for $\omega$ and $\beta$ coefficients at 1\% level. This leans in favour of the restricted PVQMLE. We also notice that the estimated coefficients and the corresponding  standard errors of the  restricted PVQMLE are fairly close to the ones obtained from the beta autoregression reported  in Table 1 of   \cite{gorgi2021beta}.

\begin{table}[h!]
	\centering
	\caption{Estimation results for the realized correlation series. Standard errors in brackets.  The bottom of the table reports the $p$-values of  the tests on the restrictions. }
	\scalebox{0.8}{
		\begin{tabular}{cccccccc}
			\cline{2-8}
			& $\hat \omega_1$ & $\hat \alpha_1$ &	$\hat \beta_1$ & $\hat \phi$ & $\hat \omega_2$ & $\hat \alpha_2$ &	$\hat \beta_2$  \\
			
			\noalign{\smallskip} \hline \noalign{\smallskip}
			
			Unrestricted & 0.01 & 0.163 & 0.822 & 22.226 & 0.055 & 0.045 & 0.898   \\
			& (0.003) & (0.013) & (0.015) & (2.745) & (0.019) & (0.007) & (0.022)  \\[0.2cm]
			Restricted  & 0.01 & 0.161  & 0.826 & 36.963  & - & - & - \\
			& (0.003) & (0.013)  & (0.015) & (1.073)  &  &  & \\
			
			\noalign{\smallskip} \hline \noalign{\smallskip}
			
			$H_0$ & $\omega_1=\omega_2$ & $\alpha_1=\alpha_2$ & $\beta_1=\beta_2$ & joint test& & &  \\
			$p$-value & 0.02 & $<$0.001 & 0.01 & $<$0.001 &&&\\
			
			\noalign{\smallskip} \hline \noalign{\smallskip}
		\end{tabular}
	}
	\label{Tab: realized corr}
\end{table}

\section{Conclusions} \label{SEC: conclusions}

We have introduced a novel methodology for the estimation of a broad range of semi-parametric time series models, where only the   conditional mean is correctly specified by a parametric function. 
Our proposed PVQMLE is based  on a Gaussian quasi-likelihood function and relies  on the specification of a parametric pseudo-variance, which does not need to be the true conditional variance of the process and it may include restrictions on parameters related to the conditional expectation. 
We have established the asymptotic properties of the PVQMLE estimator  with and without restrictions on the parameter space, and derived a test to validate the parameter restrictions. Importantly, our findings hold regardless of the correct specification of the pseudo-variance. A significant advantage of our restricted estimators is their potential to achieve greater efficiency compared to other quasi-likelihood methods found in existing literature. Additionally, our testing approach enables the development of specification tests for parametric time series models. We have demonstrated the practical application of our methodology through simulation studies and empirical cases.

\section*{Acknowledgments}

We greatly appreciate comments made by the Editor, Associate Editor and two reviewers on an earlier version of the manuscript.
Mirko Armillotta acknowledges financial support from the EU Horizon Europe programme under the Marie Skłodowska-Curie grant agreement No.101108797.

\section*{Data availability}
Codes and data to replicate the analyses in the paper are available at \url{https://github.com/mirkoarmillotta/Pseudo_variance}.

\begin{appendices}
	
\section{Proofs of results}
\label{Appendix A proof of results}
\renewcommand{\theequation}{A.\arabic{equation}}
\setcounter{equation}{0}
\setcounter{subsection}{0}

\medskip

\subsection{Proofs}

\noindent \textbf{Proof of Theorem~\ref{Thm. can}:} 
Let $L(\theta) = \E[l_t(\theta)]$ be the limit log-quasi-likelihood. In what follows we show the following intermediate results.
\begin{enumerate}[label=(\roman*)] 
	\item Uniform convergence: $\sup_{\theta \in \Theta} |{\tilde{L}_T(\theta) - L(\theta)}| \to 0$ almost surely, as $T \to \infty$. \label{Cond. uniform convergence}
	\item Identifiability: the pseudo-true parameter value $\theta_0$ is the unique maximizer of $L(\theta)$, i.e. $\E\left[ l_t(\theta) \right]   < \E\left[  l_t(\theta_0) \right] $ for all $\theta \in \Theta, \theta \neq \theta_0 $. \label{Cond. ineqaulity}
\end{enumerate}
In order to prove \ref{Cond. uniform convergence} the uniform convergence of the two summands of \eqref{triangle} should be shown.
\begin{equation} \label{triangle}
	|\tilde{L}_T(\theta) - L(\theta)| \leq |\tilde{L}_T(\theta) - L_T(\theta)| + |L_T(\theta) - L(\theta)|\,.
\end{equation}
The first term converges uniformly by Lemma~\ref{lemma1} in appendix~\ref{Appendix subsec technical}, under \ref{Ass. lower bound}-\ref{Ass. starting value}, implying that the starting value of the process is asymptotically unimportant for the quasi-likelihood contribution.
By assumption \ref{Ass. stationarity} the log-quasi-likelihood contribution $l_t(\theta)$ is stationary and ergodic. Moreover, it is uniformly bounded
\begin{align*}
	\E \, \sup_{\theta \in \Theta}\norm{ l_t(\theta) } \leq \frac{1}{2}\E \, \sup_{\gamma \in \Gamma} \norm{ \log \nu^*_t(\gamma)} + \frac{1}{2} \E \, \sup_{\theta \in \Theta}\left(  \frac{\left[ Y_t - \lambda_t(\psi) \right]^2 }{\nu^*_t(\gamma)} \right)  < \infty
\end{align*}
by assumption \ref{Ass. uniform moment}. For the continuity of the quasi-likelihood and the compactness of $\Theta$, \citet[Thm.~2.7]{straumann2006} applies providing the uniform convergence of the second term in \eqref{triangle}; in symbols $\sup_{\theta \in \Theta}|L_T(\theta) - L(\theta)|\to 0$ almost surely, as $T\to\infty$. This concludes the proof of \ref{Cond. uniform convergence}.

We now prove \ref{Cond. ineqaulity}. First note that by the uniform limit theorem $L(\theta)= \E[l_t(\theta)]$ is a continuous function and it attains at least a maximum in $\Theta$ since $\Theta$ is compact. We now prove that such maximum is unique so that it can be univocally identified. Recall that $\theta=(\psi^\prime, \gamma^{\prime})^\prime $, assumption \ref{Ass. uniform moment} provides $\E \,\, \sup_{\psi \in \Psi}\norm{l_t(\psi, \gamma)}  < \infty  $ and $\E \,\, \sup_{\gamma \in \Gamma}\norm{l_t(\psi_0, \gamma)}  < \infty  $ so also the function $l_t(\psi, \gamma)$ has at least a maximum for $\psi\in\Psi$, and $l_t(\psi_0, \gamma)$ has at least a maximum for $\gamma\in\Gamma$. Consider now $\E \left\lbrace l_t(\theta) - l_t  (\theta_0) \right\rbrace = \E \left\lbrace l_t(\theta) - l_t  (\psi_0, \gamma) \right\rbrace + \E \left\lbrace  l_t  (\psi_0, \gamma) - l_t  (\theta_0) \right\rbrace $. The first summand is bounded as follows,
\begin{align*}
	\E \left\lbrace l_t(\theta) - l_t  (\psi_0, \gamma) \right\rbrace &= \E \left\lbrace - \frac{\E [ \left( Y_t - \lambda_t(\psi) \right)^2 | \Fb_{t-1} ] }{2 \nu^*_t(\gamma)} + \frac{\nu_t}{2 \nu^*_t(\gamma)} \right\rbrace \\
	&\leq \E \left\lbrace - \frac{\nu_t}{2 \nu^*_t(\gamma)} + \frac{\nu_t}{2 \nu^*_t(\gamma)} \right\rbrace = 0  
\end{align*}
with equality if and only if  $\psi=\psi_0$ by assumption \ref{Ass. identification}. Moreover,
$\E \left\lbrace  l_t  (\psi_0, \gamma) - l_t  (\theta_0) \right\rbrace = \E \left[   l_t  (\psi_0, \gamma)  \right] - \E \left[ l_t  (\psi_0, \gamma^*) \right]  \leq 0$ by assumption \ref{Ass. unique maximizer}. 
This concludes the proof of \ref{Cond. ineqaulity}.
The consistency of the whole estimator $\hat{\theta}$ follows from \ref{Cond. uniform convergence}, \ref{Cond. ineqaulity} and the compactness of $\Theta$ by \citet[Lemma~3.1]{PoetscherandPrucha(1997)}. This implies \eqref{consistency}.

To prove the asymptotic normality of the estimator we establish additional intermediate results.
\begin{enumerate}[label=(\alph*)] 
	\item $\sqrt{T} \sup_{\theta \in \Theta} \|{S_T(\theta)- \tilde{S}_T(\theta)}\| \to 0$ almost surely, as $T \to \infty$. \label{Cond. starting value score}
	\item Define $H_T(\theta) = T^{-1} \sum_{t=1}^{T} -\partial^2 l_t(\theta)/\partial \theta \partial \theta^\prime$.  $ H_T(\theta)\to H(\theta)$ almost surely uniformly over $\theta\in\Theta$, as $T \to \infty$. \label{Cond. uniform convergence hessian}
	\item $\E[s_t(\theta_0)]=0$. \label{Cond. expected score 0}
\end{enumerate}
The condition \ref{Cond. starting value score} is satisfied by Lemma~\ref{lemma2} in appendix~\ref{Appendix subsec technical}, under \ref{Ass. lower bound}-\ref{Ass. starting value} and \ref{Ass. starting value derivative} implying that initial values of the process do not affect the asymptotic distribution of the PVQMLE.

Consider the second derivative of the log-quasi-likelihood contribution.
\begin{align}
	\frac{\partial^2 l_t(\theta)}{\partial \theta \partial \theta^\prime} &= \left( \frac{1}{2 \nu^{*2}_t(\gamma)} - \frac{[ Y_t - \lambda_t(\psi)]^2}{\nu^{*3}_t(\gamma)} \right) \frac{\partial \nu^*_t(\gamma)}{\partial \theta} \frac{\partial \nu^*_t(\gamma)}{\partial \theta^\prime} \label{second derivative}\\
	& \quad \quad - \frac{ Y_t - \lambda_t(\psi)}{\nu^{*2}_t(\gamma)} \left( \frac{\partial \lambda_t(\psi)}{\partial \theta} \frac{\partial \nu^*_t(\gamma)}{\partial \theta^\prime} - \frac{\partial \nu^*_t(\gamma)}{\partial \theta} \frac{\partial \lambda_t(\psi)}{\partial \theta^\prime} \right)  \nonumber \\
	& \quad \quad - \frac{1}{\nu^*_t(\gamma)} \frac{\partial \lambda_t(\psi)}{\partial \theta} \frac{\partial \lambda_t(\psi)}{\partial \theta^\prime} + \frac{ Y_t - \lambda_t(\psi)}{\nu^*_t(\gamma)} \frac{\partial^2 \lambda_t(\psi)}{\partial \theta \partial \theta^\prime} \nonumber \\
	& \quad \quad + \left( \frac{[ Y_t - \lambda_t(\psi)]^2}{2\nu^{*2}_t(\gamma)} - \frac{1}{2\nu^*_t(\gamma)} \right)  \frac{\partial^2 \nu^*_t(\gamma)}{\partial \theta \partial \theta^\prime}\,. \nonumber
\end{align}
Assumption  \ref{Ass. moment second derivative} and the Cauchy-Schwarz inequality yield $\E \, \sup_{\theta \in \Theta} \norm{\partial^2 l_t(\theta)/\partial \theta_i \partial \theta_j}   < \infty$ for all $i,j=1,\dots,m$. 
 Furthermore, the second derivative is a continuous, stationary and ergodic sequence. Then, an application of \citet[Thm.~2.7]{straumann2006} provides the condition \ref{Cond. uniform convergence hessian}. Note that since in this case $\partial \lambda_t(\psi)/\partial \gamma=\partial \nu^*_t(\gamma)/\partial \psi=0$ the matrix $H(\theta_0)$ is block diagonal with diagonal block matrices $H_\psi(\theta_0)= \E \left[ - \partial^2 l_t(\theta_0)/\partial \psi \partial \psi^\prime \right] $ and $H_{\gamma}(\theta_0) = \E \left[ - \partial^2 l_t(\theta_0)/\partial \gamma \partial \gamma^{\prime} \right] $. The former is defined in \eqref{hessians uncons}.


For establishing the asymptotic normality of the estimator $\hat{\theta}$ the proof of \ref{Cond. expected score 0} is needed. Let $s_t(\theta_0)=[s_t^{(\psi)}(\theta_0)^\prime, s_t^{(\gamma)}(\theta_0)^\prime ]^\prime $ be the partition of the score between mean and pseudo-variance parameters. Observe that $\E( s_t^{(\psi)}(\theta_0) | \Fb_{t-1} ) = 0$ but $\E\left( s_t(\theta_0) | \Fb_{t-1} \right)  \neq 0$. Note that $\sup_{\theta \in \Theta} \norm{\partial l_t(\theta)/\partial \theta_i}\leq 2\left[  \sup_{\theta \in \Theta} \norm{ l_t(\theta)}\right]^{1/2} \left[ \sup_{\theta \in \Theta} \norm{\partial^2 l_t(\theta)/\partial \theta_i \partial \theta_i}\right]^{1/2}$, by  \citet[p.~115]{rudin1976principles}. Moreover, $\E \, \sup_{\theta \in \Theta} \norm{ l_t(\theta)}   < \infty$, and 
$\E \, \sup_{\theta \in \Theta} \norm{\partial^2 l_t(\theta)/\partial \theta_i \partial \theta_j}   < \infty$. Then an application of Cauchy-Schwarz inequality entails $\E \, \sup_{\theta \in \Theta} \norm{\partial l_t(\theta)/\partial \theta_i}   < \infty$. Finally, $\lnorm{\partial l_t(\theta)/\partial \theta} \leq \sup_{\theta \in \Theta} \lnorm{\partial l_t(\theta)/\partial \theta}$ and an application of the dominated convergence theorem leads to $\E\left[ \partial l_t(\theta)/\partial \theta\right] = \partial  \E\left[ l_t(\theta)\right]/\partial \theta $. By noting that $\theta_0$ is the unique maximizer of $\E\left[ l_t(\theta)\right]$ the result \ref{Cond. expected score 0} follows.

Using the formula \eqref{score} some tedious algebra allows to show that  $\E \lnorm{s_t(\theta_0)s_t(\theta_0)'} < \infty$, by  \ref{Ass. moment second derivative} and an application of Cauchy-Schwarz inequality. Therefore $I(\theta_0)= \E \left[s_t(\theta_0)s_t(\theta_0)^\prime \right]$ is finite.

For $T$ large enough $\hat{\theta} \in \dot{\Theta}$ by \ref{Ass. interior}, so the following derivatives exist almost surely
\begin{align*}
	0 = \sqrt{T}\tilde{S}_T(\hat{\theta}) = \sqrt{T}S_T(\hat{\theta}) + o_p(1) = \sqrt{T}S_T(\theta_0) - H_T(\bar{\theta})\sqrt{T}( \hat{\theta} - \theta_0)   + o_p(1),
\end{align*}
where the first equality comes from the definition \eqref{quasi likelihood}, the second equality holds by \ref{Cond. starting value score}, and the third equality is obtained by Taylor expansion at $\theta_0$ with $\bar{\theta}$ lying between $\hat{\theta}$ and $ \theta_0$. By assumption \ref{Ass. score clt} and \ref{Cond. expected score 0} we have $\sqrt{T}S_T(\theta_0)  \xrightarrow{d} N(0, I(\theta_0))$. This fact and \ref{Cond. uniform convergence hessian} establish the asymptotic normality of the estimator $\hat{\theta}$ with covariance matrix $\Sigma(\theta_0)=H^{-1}(\theta_0) I(\theta_0) H^{-1}(\theta_0)$  by assumption \ref{Ass. hessians}, where
\begin{equation} \label{block matrices}
	H(\theta_0)=
	\begin{pmatrix}
		H_\psi(\theta_0) & 0 \\
		0 & H_\gamma(\theta_0)
	\end{pmatrix}\,,\quad
	I(\theta_0)=
	\begin{pmatrix}
		I_\psi(\theta_0) & I_{\psi,\gamma}(\theta_0) \\
		I_{\psi,\gamma}(\theta_0)' & I_\gamma(\theta_0)
	\end{pmatrix}\,,
\end{equation}
with $H_x(\theta_0) = \E \left[ - \partial^2 l_t(\theta_0)/\partial x \partial x^{\prime} \right] $,  $I_x(\theta_0)= \E [s_t^{(x)}(\theta_0)s_t^{(x)}(\theta_0)^\prime ]$ and $I_{x,z}(\theta_0)= \break \E [s_t^{(x)}(\theta_0)s_t^{(z)}(\theta_0)^\prime]$. In particular, standard algebra shows that $I_\psi(\theta_0)$ equals \eqref{hessians uncons}. See also equation \eqref{sqmle mean}. A suitable block matrix multiplication of \eqref{block matrices} provides 
\begin{equation*}
	\Sigma(\theta_0)=
	\begin{pmatrix}
		\Sigma_\psi(\theta_0) & \Sigma_{\psi,\gamma}(\theta_0) \\
		\Sigma_{\psi,\gamma}(\theta_0)' & \Sigma_\gamma(\theta_0)
	\end{pmatrix}\,,
\end{equation*}
where $\Sigma_\psi(\theta_0)$ takes the form defined in \eqref{asymp. normality}. In addition, note that for the marginal property of the multivariate Gaussian distribution result \eqref{asymp. normality} holds with covariance matrix $\Sigma_\psi$ being the partition of $\Sigma(\theta_0)$ for the mean parameters $\psi$.

The positive definiteness of the matrix $\Sigma(\theta_0)$ follows since for all $\delta \in \R^{m}$, with $\delta \neq 0$, we have $H(\theta_0)^{-1}\delta \neq 0$ as $H(\theta_0)^{-1}$ is full rank by \ref{Ass. hessians}. Now by setting $\eta = H(\theta_0)^{-1}\delta$ we have that $\eta'I(\theta_0)\eta >0$ by \ref{Ass. hessians}. Therefore, it follows that $\delta'  H(\theta_0)^{-1} I(\theta_0) H(\theta_0)^{-1} \delta >0$. The principal submatrices of $\Sigma(\theta_0)$ are also positive definite. \hfill$\square$\\ 

\noindent \textbf{Proof of Corollary~\ref{Cor. correct variance}:} 
Condition \ref{Ass. score clt} is not required since in this case is easily showed by \eqref{score} that $\E\left( s_t(\theta_0) | \Fb_{t-1} \right) = 0$. 
Recall that $\sqrt{T} s_T(\theta_0)= T^{-1/2}\sum_{t=1}^{T} U_t$ where $ U_t = s_t(\theta_0)$. Note that $\left\lbrace U_t, \Fb_t \right\rbrace $ is a stationary martingale difference, and due to \ref{Ass. moment second derivative}-\ref{Ass. hessians} it has a finite and positive definite second moments matrix. Then \ref{Ass. score clt} follows by the central limit theorem for martingales \cite[]{billingsley1961cltmds} and the Cramér-Wold device. The consistency and asymptotic normality of $\hat{\theta}$ follow as above.
Finally, in view of \eqref{sqmle mean} and $\E( s_t^{(\psi)}(\theta_0) | \Fb_{t-1}) = 0$
\begin{equation*}
	\mathrm{Var} \big[ H_\psi^{-1}(\theta_0) s_t^{(\psi)}(\theta_0) - I_\psi^{-1}(\theta_0) s_t^{(\psi)}(\theta_0) \big] = \Sigma_\psi - I_\psi
\end{equation*}
being necessarily positive semi-definite. \hfill$\square$\\ 

\noindent \textbf{Proof of Corollary~\ref{Cor. set consistency}:} Analogously to the proof of Theorem~\ref{Thm. can}, \ref{Ass. stationarity}-\ref{Ass. starting value}  guarantee that $L_t(\theta)$ is continuous and a.s.~uniformly convergent to $\E[l_t(\theta)]$. By recalling that $\Theta$ is compact the result follows by \citet[Lemma~4.2]{PoetscherandPrucha(1997)}. \hfill$\square$\\

\noindent \textbf{Proof of Theorem~\ref{Thm. constrained can}:}
The consistency of $\hat \theta_R$ follows  from the fact that by the proof of Theorem~\ref{Thm. can}  we have that $\E  [l_t(\psi,\gamma)] \le \E [ l_t(\psi_0,\gamma^*)]$ for any $\theta\in\Theta$ with equality holding only if  $\theta=({\psi_0}',{\gamma^*}')'$, and assumption \ref{Ass. constrain holds} ensures that $({\psi_0}',{\gamma^*}')'\in \Theta_R$ with $\Theta_R \subseteq \Theta$. The consistency in \eqref{consistency cons} follows. The asymptotic normality of the estimator $\hat \theta_R$ follows as in the proof of Theorem  \ref{Thm. can} with covariance matrix  $\Sigma(\theta_0)=H^{-1}(\theta_0) I(\theta_0) H^{-1}(\theta_0)$. In this case Hessian and Fisher information matrices can be written in the following block matrix form
\begin{equation} \label{block matrices constr}
	H(\theta_0)=
	\begin{pmatrix}
		H_\psi(\theta_0) & H_{\psi,\gamma_2}(\theta_0) \\
		H_{\psi,\gamma_2}(\theta_0)' & H_{\gamma_2}(\theta_0)
	\end{pmatrix}\,,\quad
	I(\theta_0)=
	\begin{pmatrix}
		I_\psi(\theta_0) & I_{\psi,\gamma_2}(\theta_0) \\
		I_{\psi,\gamma_2}(\theta_0)' & I_{\gamma_2}(\theta_0)
	\end{pmatrix}\,.
\end{equation}
Moreover, recall that 
\begin{equation} \label{inverse hessian constr}
	H^{-1}(\theta_0)=D(\theta_0)=
	\begin{pmatrix}
		D_\psi(\theta_0) & D_{\psi,\gamma_2}(\theta_0) \\
		D_{\psi,\gamma_2}(\theta_0)' & D_{\gamma_2}(\theta_0)
	\end{pmatrix}\,.
\end{equation}
By computing $\Sigma(\theta_0)$ using the block matrix multiplication as defined in \eqref{block matrices constr} and \eqref{inverse hessian constr} the partition of $\Sigma(\theta_0)$ for the mean parameters $\psi$ equals $\Sigma_R$. This entails \eqref{asymp. normality cons}. \hfill$\square$\\ 

\noindent \textbf{Proof of Theorem~\ref{Thm. stationarity inar}:}
The result follows by a combination of \citet[Thm.~1-2]{doukhan_2012} and the results of \citet[Sec.~4.1]{doukhan_2012} given that $X_j \sim D_X(a,b)$ and $\E(X_j)=a$. Then the process is stationary, ergodic and $\E(Y_t) < \infty$. The same results show that the process is  $\beta$-mixing with geometrically decaying coefficients.  Finally, following  \citet[Sec.~3]{lat1997} we conclude that $\E(Y_t^2) < \infty$. \hfill$\square$\\

\noindent \textbf{Proof of Theorem~\ref{Thm. can inar}:}
To prove the results we have to prove conditions~\ref{Ass. stationarity}-\ref{Ass. constrain holds} for the specified model.
First note that since the pseudo-variance $\nu^*_t$ defined in \eqref{inar pseudo-var} is correctly specified we have that $\nu^*_t(\cdot)=\nu_t(\cdot)$. Moreover, \ref{Ass. constrain holds} holds.
The condition \ref{Ass. stationarity} holds by Theorem~\ref{Thm. stationarity inar}. \ref{Ass. lower bound} holds since a.s. $\nu_t(\gamma) \geq \omega_2$. Note that a.s. $\sup_{\gamma \in \Gamma} |\log \nu_t(\gamma) | \leq \sup_{\gamma \in \Gamma} (\nu_t(\gamma)+1)/\min\{\omega_2,1\}$ and $\sup_{\theta \in \Theta} (Y_t-\lambda_t(\psi))^2/\nu_t(\gamma) \leq (2 Y_t^2 + 2 \sup_{\psi \in \Psi}\lambda_t^2(\psi))/\omega_2$. By the $c_p$ inequality it holds that $\E \sup_{\psi \in \Psi} \lambda_t^r(\psi) < \infty$ and $\E \sup_{\gamma \in \Gamma} \nu_t^r(\gamma) < \infty$ for $r \leq 2$ so the moments in \ref{Ass. uniform moment} are finite.

We prove $\ref{Ass. identification}$ by contradiction. We have that a.s. $\lambda_t(\psi) -  \lambda_t(\psi_0)  = \omega_1 - \omega_{1,0} + (a - a_{0}) Y_{t-1}$. If $\lambda_t(\psi)= \lambda_t(\psi_0)$ a.s. with $\omega_1 \neq \omega_{1,0}$ then $0 \neq \omega_{1,0} - \omega_1 = (a - a_{0}) Y_{t-1}$ a.s. and the equality will be possible only if $(a - a_{0}) \neq 0$ and $Y_{t-1}$ equals a.s. a non-zero constant. However, $Y_{t-1}$ is non-constant. Therefore, if $\lambda_t(\psi)= \lambda_t(\psi_0)$ a.s. then $\omega_1 = \omega_{1,0}$ and $0 = (a - a_{0}) Y_{t-1}$. Now to have  $\lambda_t(\psi)= \lambda_t(\psi_0)$ a.s. with $a \neq a_{0}$ we shall have that $Y_{t-1}=0$ a.s. but this is impossible since $Y_{t-1}$ is non-constant. Hence, if $\lambda_t(\psi)= \lambda_t(\psi_0)$ a.s. then $a = a_{0}$. Analogous results hold for $\nu_t(\gamma)$.

Assumptions \ref{Ass. starting value}-\ref{Ass. starting value derivative} are trivially satisfied here since $\lambda_t$ and $\nu_t$ are initialized  using the first observation of the sample so $\lambda_t(\cdot)= \tilde \lambda_t(\cdot)$ and $\nu_t(\cdot)= \tilde \nu_t(\cdot)$. Condition \ref{Ass. unique maximizer} is verified since $\gamma^*$ is the true parameter vector of the variance, say $\gamma_0$, so $\nu_t(\gamma_0)=\nu_t$ a.s. and 
	\begin{equation*}
		\E\left[ l_t(\psi_0,\gamma) - l_t(\psi_0,\gamma_0)\right] = \E\left[ \frac{1}{2}\log \frac{\nu_t}{\nu_t(\gamma)}+\frac{1}{2}-\frac{\nu_t}{2 \nu_t(\gamma)} \right]  \leq \E \left[\frac{\nu_t}{2 \nu_t(\gamma)} - \frac{1}{2} +\frac{1}{2}-\frac{\nu_t}{2 \nu_t(\gamma)} \right] =0 
	\end{equation*}
where the inequality follows by $\log(x) \leq x-1$ for $x>0$. So $\E\left[ l_t(\psi_0,\gamma) \right]  \leq  \E \left[ l_t(\psi_0,\gamma_0)\right]$ with equality if and only if (henceforth, iff) $\nu_t(\gamma)=\nu_t(\gamma_0)$ a.s. but by \ref{Ass. identification} this happens iff $\gamma=\gamma_0$. Therefore $\gamma_0$ is unique maximizer of \eqref{pseudo_true_par}. Let $0_k$ be a $k \times 1$ vector of zeros.
Recall that $\partial \lambda_t(\psi)/\partial \theta=(1, Y_{t-1},0_2')'=\bar Y_{t-1}$ and $\partial \nu_t(\gamma)/\partial \theta$ is a permutation of the elements of $\bar Y_{t-1}$. Therefore an application of H\"{o}lder's inequality and $\E(Y_t^8)< \infty$ provide \ref{Ass. moment second derivative}. To prove \ref{Ass. hessians} note that the elements of $\partial \lambda_t(\psi)/\partial \psi=\partial \nu_t(\gamma)/\partial \gamma=(1, Y_{t-1})'$ are linearly independent and $\nu^*_t(\gamma^*)=\nu_t$ so by employing the results of Lemma~\ref{Lem invertibility} in appendix~\ref{Appendix subsec technical} the sufficient condition \ref{Ass. linear indipendence} holds. The same follows for the restricted estimators since $\partial \nu_t(\gamma)/\partial \gamma_2$ is a subvector of $(1, Y_{t-1})'$.

	Finally, recall that the process $\left\lbrace Y_t\right\rbrace $ is $\beta$-mixing with coefficients $\beta(n) \leq C \rho^n$ where $C,\rho$ are positive constants and $\rho \in (0,1)$. Following \citet[Sec.~A.3]{fran2019} the score contribution $s_t(\theta_0)$ is also $\beta$-mixing with coefficients $\beta_s(n) \leq \beta (n-1)$ for $n \geq 1$. By recalling that $\alpha_s(n) \leq \beta_s(n)$ for $n \geq 1$ and $\alpha_s(0) \leq 1/4$, we have that $\sum_{n=0}^{\infty}[\alpha(n)]^{\delta/(2+\delta)} < \infty $ for some $\delta >0$. Moreover, by  a combination of H\"{o}lder's and $c_p$ inequalities, $\E(Y_t)^8 < \infty$ is sufficient to show that $\E(\eta's_t(\theta_0))^{2+\delta} < \infty$ for $\delta=2$ and for all $\eta \in \R^m$ with $\eta \neq 0$. Therefore, an application of the Cramér-Wold device and the central limit theorem for $\alpha$-mixing processes \cite[Thm.~A.4]{fran2019} shows that $\sqrt{T}S_T(\theta_0) \xrightarrow{d} N(0, I(\theta_0))$ as $T  \to \infty$. This proves \ref{Ass. score clt}.
\hfill$\square$\\

\noindent \textbf{Proof of Theorem~\ref{Thm. can beta}:}
Since the observations are generated from a beta distribution, the results of Theorem~\ref{Thm. stationarity beta} guarantee that \ref{Ass. stationarity} and the restrictions of \ref{Ass. constrain holds} are satisfied with $\nu^*_t(\cdot)=\nu_t(\cdot)$.  Define $\delta_i = \omega_i+\alpha_i+\beta_i$ for $i=1,2$.
 The restrictions on the parameter space imply that a.s. $0 < \omega_1 \leq \lambda_t(\psi) \leq \delta_1 <1$, $0 < \omega_2 \leq \mu_t(\gamma) \leq \delta_2 <1$ and $0 < \underline \nu \leq \nu_t(\gamma) < \bar \nu <1$ for any $\theta \in \Theta $ and any $t\geq1$ where $\bar \nu =1/(1+\phi)<1$ since $\mu_t(1-\mu_t) < 1$ and $\underline \nu = \min\left\lbrace \underline{\nu}_1,\underline{\nu}_2\right\rbrace $ where $\underline{\nu}_1 = \omega_2(1-\omega_2)/(1+\phi)$ and $\underline{\nu}_2 = \delta_2(1-\delta_2)/(1+\phi)$. All these processes are a.s. bounded in the $(0,1)$ interval for any $\theta \in \Theta$ therefore all their sup-moments are bounded. Hence, \ref{Ass. uniform moment} and \ref{Ass. lower bound} hold.

$\ref{Ass. identification}$ is proved by contradiction. Assume that a.s. $\lambda_{t-1}(\psi)= \lambda_{t-1}(\psi_0) = \lambda_{t-1}$. Then $\lambda_t(\psi) -  \lambda_t(\psi_0)  = \omega_1 - \omega_{1,0} + (\alpha_1 - \alpha_{1,0}) Y_{t-1} + (\beta_1 - \beta_{1,0}) \lambda_{t-1}$. If $\lambda_t(\psi)= \lambda_t(\psi_0)$ a.s. with $\omega_1 \neq \omega_{1,0}$ then $0 \neq \omega_{1,0} - \omega_1 = (\alpha_1 - \alpha_{1,0}) Y_{t-1} + (\beta_1 - \beta_{1,0}) \lambda_{t-1}$ a.s. and the equality will be possible only if $(\alpha_1 - \alpha_{1,0}) \neq 0$ and $Y_{t-1}$ equals a non-zero constant a.s. and/or  $(\beta_1 - \beta_{1,0}) \neq 0$ and $\lambda_{t-1}$ equals a non-zero constant. However, $Y_{t-1}$ is non-constant and since $\alpha_1>0$ this is true also for $\lambda_{t-1}$. Therefore, if $\lambda_t(\psi)= \lambda_t(\psi_0)$ a.s. then $\omega_1 = \omega_{1,0}$ and $0 = (\alpha_1 - \alpha_{1,0}) Y_{t-1} + (\beta_1 - \beta_{1,0}) \lambda_{t-1}$. Now to have  $\lambda_t(\psi)= \lambda_t(\psi_0)$ a.s. with $\alpha_1 \neq \alpha_{1,0}$ and $\beta_1 \neq \beta_{1,0}$ we shall have that a.s. $Y_{t-1}=\lambda_{t-1}=0$ but this is impossible since $Y_{t-1}$ and $\lambda_{t-1}$ are non-constant. Therefore, if $\lambda_t(\psi)= \lambda_t(\psi_0)$ a.s. then $\alpha_1 = \alpha_{1,0}$ and $\beta_1 = \beta_{1,0}$. An analogous result holds for $\mu_t(\gamma)$, consequently $\nu_t(\gamma)=\nu_t(\gamma_0)$ a.s. if and only if $\gamma=\gamma_0$. Then, \ref{Ass. unique maximizer} holds following the same arguments provided in the proof of Theorem~\ref{Thm. can inar}.

Recall that $\tilde Y_{-i} \in [0,1]$ for $i=0,1,\dots$ so a.s. $|\lambda_t(\psi) - \tilde \lambda_t(\psi) | = \beta_1^t | \lambda_0(\psi) - \tilde \lambda_0(\psi)|\leq 2\beta_1^t$, $|\mu_t(\gamma) - \tilde \mu_t(\gamma) | \leq 2\beta_2^t$. The variance is a function of $\mu_t$ so in simplified notation $\partial \nu_t(\gamma, \mu)/\partial \mu  = (1-2\mu)/(1+\phi)$ and $|1-2\mu| \leq c < 1$ since $0 < \mu <1$, therefore by the mean value theorem a.s. $|\nu_t(\gamma) - \tilde \nu_t(\gamma) | \leq 2c\beta_2^t$. This implies that,
as $t \to \infty$, $a_t,b_t \to 0$ e.a.s. where $e.a.s.$ means \textit{exponentially fast a.s. convergence} \cite[Sec.~2.1]{straumann2006}. Then, the limits in \ref{Ass. starting value} converge e.a.s to 0.

Recall that $0_k$ is a $k \times 1$ vector of zeros.
Define $ Z_t(\theta)= (1, Y_t, \lambda_t(\psi),0_4')'$, $C_1=\| (1,1,1,0_4')'\|$ and $\sup_{\psi \in \Psi}\beta_1=\rho_1$. Note that 
\begin{equation*}
\frac{\partial \lambda_t(\psi)}{\partial \theta} = Z_{t-1}(\theta) + \beta_1 \frac{\partial \lambda_{t-1}(\psi)}{\partial \theta}\,, ~~~~~~ \frac{\partial \nu_t(\gamma)}{\partial \theta} =\frac{1-2\mu_t(\gamma)}{1+\phi}\frac{\partial \mu_t(\gamma)}{\partial \theta}+\frac{\mu_t^2(\gamma)-\mu_t(\gamma)}{(1+\phi)^2}1=A(\theta)+B(\theta)\,,
\end{equation*}
and $c_t \to 0$ e.a.s. by \citet[Lem.~A.2]{gorgi2021beta}. Then, for $t$ large enough, with probability 1
\begin{equation*}
	\sup_{\theta \in \Theta} \lnorm{\frac{\partial \lambda_t(\psi)}{\partial \theta}}  \leq  \sum_{i=0}^{t-1}\rho_1^i \sup_{\theta \in \Theta}\lnorm{Z_{t-1-i}(\theta)} + \rho_1^t \sup_{\theta \in \Theta} \frac{\partial \lambda_0(\psi)}{\partial \theta} \leq  C_1\sum_{i=0}^{\infty}\rho_1^i + 1 = M < \infty\,,
\end{equation*}
since $\rho_1 <1$. By similar arguments $\sup_{\theta \in \Theta}  \lnorm{\partial \mu_t(\gamma)/\partial \theta}  \leq K $  and $\sup_{\theta \in \Theta}  \lnorm{\partial \nu_t(\gamma)/\partial \theta}  \leq 3K+2$, for $t$ large enough, where $K$ is a positive constant. 
By employing again the mean value theorem it follows that, for $t$ large enough and with probability 1
\begin{equation*}
	d_t \leq \sup_{\theta \in \Theta} \sup_{\mu \in (0,1)} \lnorm{\frac{\partial}{\partial \theta}\left( \frac{ \partial \nu_t(\gamma, \mu)}{\partial \mu}\right)} \sup_{\gamma \in \Gamma}|\mu_t(\gamma)-\tilde \mu_t(\gamma)| \leq (3+2K)\sup_{\gamma \in \Gamma}|\mu_t(\gamma)-\tilde \mu_t(\gamma)|
\end{equation*}
 converging to 0 e.a.s. as $t \to\infty$.
Then, the limits in \ref{Ass. starting value derivative} converge e.a.s to 0 and are of order $\mathcal{O}(t^{-\delta})$.

Recall that $O_{m,n}$ is a $m \times n$ matrix of zeros. The second derivative has the form 
\begin{equation*}
	\frac{\partial^2 \lambda_t(\psi)}{\partial \theta \partial \theta'} = \dot Z_{t-1}(\theta) + \beta_1 \frac{\partial^2 \lambda_{t-1}(\psi)}{\partial \theta \partial \theta'}\,, ~~~~~
	\dot Z_{t-1}(\theta) = 
	\begin{pmatrix}
		O_{2,3} & 0_2 \\
		\frac{\partial \lambda_{t-1}(\psi)}{\partial \theta'} & 0\\
		0_3' & 0
	\end{pmatrix}\,.
\end{equation*}	
Following the same arguments of the first derivative, for $t$ large enough and probability 1
\begin{align*}
	\sup_{\theta \in \Theta} \lnorm{\frac{\partial^2 \lambda_t(\psi)}{\partial \theta \partial \theta'}} \leq  \sum_{i=0}^{\infty}\rho_1^i \sup_{\theta \in \Theta}\lnorm{\dot Z_{t-1-i}(\theta)} + 1 \leq \frac{C_2 M}{1-\rho_1}+1 < \infty\,,
\end{align*}
where $C_2$ is a positive constant depending on the type of matrix norm $\| \cdot \|$ employed.
Analogously, $\sup_{\theta \in \Theta} \lnorm{\partial^2 \mu_t(\gamma)/\partial \theta \partial \theta'}$ and $\sup_{\theta \in \Theta} \lnorm{\partial^2 \nu_t(\gamma)/\partial \theta \partial \theta'}$ are a.s. bounded by a constant  so \ref{Ass. moment second derivative} is verified.

Consider a deterministic vector $\eta \in \R^m$ with $\eta=(\eta_1',\eta_2)'$ where $\eta_1$ is of dimension 3 and $\eta_2$ is a scalar. By appealing the results of Lemma~\ref{Lem invertibility} in appendix~\ref{Appendix subsec technical}, we prove \ref{Ass. hessians} by showing that $\eta_1'\partial \lambda_t(\psi_0)/\partial \psi=0$ a.s.
if and only if (henceforth, iff) $\eta_1=0$. The proof is by contradiction. Assume that $\eta_1'\partial \lambda_t(\psi_0)/\partial \psi=0$ a.s. for some $\eta_1 \neq 0$. Then $\eta_1'\partial \lambda_{t-1}(\psi_0)/\partial \psi=0$ a.s. by stationarity. Therefore from the formula of the first derivative we should have $\eta_1' Z_{t-1}(\psi_0)=\eta_1'(1,Y_{t-1},\lambda_{t-1}(\psi_0))'=0$ a.s. for some $\eta_1 \neq 0$. However, this is impossible since  $ Z_{t-1}(\psi_0)$ 
has linearly independent elements so it follows that $\eta_1' Z_{t-1}(\psi_0)=0$ a.s.
iff $\eta_1=0$. 
Recall that $\gamma=(\gamma_1, \phi)'$ where $\gamma_1=(\omega_2,\alpha_2,\beta_2)'$.  Note that
$$
\eta'\frac{\partial \nu_t(\gamma_0)}{\partial \gamma}=\eta_1'\frac{\partial \nu_t(\gamma_0)}{\partial \gamma_1}+\eta_2\frac{\partial \nu_t(\gamma_0)}{\partial \phi}=\eta_1'\frac{1-2\mu_t(\gamma_0)}{1+\phi}\frac{\partial \mu_t(\gamma_0)}{\partial \gamma_1}- \eta_2 \frac{\mu_t(\gamma_0)(1-\mu_t(\gamma_0))}{(1+\phi)^2}=\eta_1'm_tn_t-\eta_2o_t
$$
with obvious notation.
We appeal again the proof by contradiction so assume that $\eta'\partial \nu_t(\gamma_0)/\partial \gamma=0$ a.s. for some $\eta\neq 0$. We consider three cases. (i) $\eta_1\neq 0, \eta_2=0$.
We have that $m_t \neq 0$ a.s. since $\mu_t(\gamma_0)$ is non-degenerate, therefore it should be that $\eta_1'n_t=0$ a.s. for some $\eta_1 \neq 0$, however $n_t$ has linearly independent elements, following the same arguments of $\eta_1'\partial \lambda_t(\psi_0)/\partial \psi$ above, so the assumed statement cannot be true.  (ii) $\eta_1 = 0, \eta_2 \neq 0$. In this case we have $o_t >0$ a.s., by definition, so $\eta_2 o_t=0$ a.s. cannot occur since $\eta_2 \neq 0$. (iii) $\eta_1 \neq0, \eta_2 \neq 0$. In this case we shall have a.s. $\eta_1' n_t = \eta_2 m_t^{-1}o_t$ and therefore $\beta_2 \eta_1' n_{t-1} = \eta_2 m_t^{-1}o_t - \eta_1'Z_{t-1}(\gamma_0)$ where $Z_{t-1}(\gamma_0)=(1,Y_{t-1},\mu_{t-1}(\gamma_0))'$. However this cannot hold because the left-hand side is $\Fb_{t-2}$-measurable whereas the right-hand side is not since it depends on $Y_{t-1}$.
Then $\eta'\partial \nu_t(\gamma_0)/\partial \gamma = 0$ a.s. iff $\eta=0$.
Therefore \ref{Ass. linear indipendence} holds and \ref{Ass. hessians} follows.
Noting that $\gamma_2=\phi$,  condition \ref{Ass. linear indipendence} holds by the arguments in (ii) so \ref{Ass. hessians} holds also for the restricted estimator.

Finally, \ref{Ass. score clt} holds as in the proof of Corollary~\ref{Cor. correct variance} because the score contribution $s_t(\theta_0)$ is a martingale difference sequence and therefore $\sqrt{T}S_T(\theta_0) \xrightarrow{d} N(0, I(\theta_0))$ as $T  \to \infty$.
\hfill$\square$\\

\noindent \textbf{Proof of Proposition~\ref{Thm. normal eff}:}
Under the conditions of Proposition~\ref{Thm. normal eff}, $p=k=1$ so $\psi$ and $\gamma$ are scalar. In particular, $\gamma=\gamma_1 \in \R$, i.e. there are no free nuisance parameters $\gamma_2$. So, under the results of Theorem~\ref{Thm. constrained can}, following the notation for restricted estimators defined below assumption~\ref{Ass. constrain holds}, it is not hard to show that the limiting covariance of the restricted estimator is a scalar and takes the form $\Sigma_R=H^{-1}_\psi(\theta_0)I_\psi(\theta_0)H^{-1}_\psi(\theta_0)$ with 
\begin{equation} \label{hessian}
	H_\psi(\theta_0)  = \E\left[ 
	\frac{1}{\nu_t(\gamma_0)} \frac{\partial \lambda_t(\psi_0)}{\partial \psi}^2 
	+
	\frac{1}{2 \nu^2_t(\gamma_0)} \frac{\partial \nu_t(\gamma_0)}{\partial \psi}^2 
	\right], 
\end{equation}
\begin{align} \label{fisher matrix}
	I_\psi(\theta_0) = & \, \, \E\left[ 
	\frac{1}{\nu_t(\gamma_0)} \frac{\partial \lambda_t(\psi_0)}{\partial \psi}^2 
	+ 
	\frac{h_t}{2 \nu^3_t(\gamma_0)} 
	\left(   \frac{\partial \lambda_t(\psi_0)}{\partial \psi} \frac{\partial \nu_t(\gamma_0)}{\partial \psi} +  \frac{\partial \nu_t(\gamma_0)}{\partial \psi} \frac{\partial \lambda_t(\psi_0)}{\partial \psi} \right) \right]
	\nonumber \\
	& 
	+ \E\left[ \left( \frac{k_t}{ \nu^2_t(\gamma_0)} - 1 \right) \frac{1}{4 \nu^2_t(\gamma_0)} \frac{\partial \nu_t(\gamma_0)}{\partial \psi}^2 
	\right],
\end{align}
where $h_t= \E[ \left( Y_t - \lambda_{t}(\psi_0) \right)^3 | \Fb_{t-1} ]$  and $k_t= \E[ \left( Y_t - \lambda_{t}(\psi_0) \right)^4 | \Fb_{t-1} ]$ by \ref{Ass. conditional moments}. By Corollary~\ref{Cor. correct variance}, the limiting covariance of the unrestricted estimator, $I_\psi^{-1}$, is the reciprocal  expected value of the first summand of \eqref{hessian}.  
	In the case \ref{Ass. efficiency}\textbf{.a} we have that $h_t=0$ and $k_t \leq 3 \nu_t(\gamma_0)$, with equality if and only if $q(\cdot)$ is Gaussian. Hence, from \eqref{fisher matrix} $I_\psi(\theta_0) \leq H_\psi(\theta_0)$ and $\Sigma_R \leq H_\psi^{-1}(\theta_0) \leq I_\psi^{-1}$ where the last inequality holds since the second summand in \eqref{hessian} is non-negative. In the case \ref{Ass. efficiency}\textbf{.b} we have that $h_t > 0$ and $\partial \lambda_t(\psi_0)/\partial \psi \,\, \partial \nu_t(\gamma_0)/\partial \psi <0$ or $h_t < 0$ and $\partial \lambda_t(\psi_0)/\partial \psi \,\, \partial \nu_t(\gamma_0)/\partial \psi >0$. In both scenarios the second summand in \eqref{fisher matrix} is negative so $I_\psi(\theta_0) < H_\psi(\theta_0)$. The result follows as above. \hfill$\square$\\


\subsection{Technical lemmas}
\label{Appendix subsec technical}
\begin{lemma}
	\label{lemma1}
	Consider the PVQMLE in  \eqref{unconstrained} with log-quasi-likelihood \eqref{quasi likelihood}. Under conditions~\ref{Ass. lower bound}-\ref{Ass. starting value}, almost surely as $T \to \infty$,
	$\sup_{\theta \in \Theta} |\tilde{L}_T(\theta) - L_T(\theta)| \to 0$.

\end{lemma}

\noindent \textbf{Proof of Lemma~\ref{lemma1}:}
From assumption  \ref{Ass. lower bound}, we have that
\begin{align*}
	& \quad \, \sup_{\theta \in \Theta} |{l_t(\theta) - \tilde{l}_t(\theta)}| \\
	&\leq \sup_{\theta \in \Theta} \norm{\frac{[\tilde{\lambda}_t(\psi) - \lambda_t(\psi) ] [  \tilde{\lambda}_t(\psi)+\lambda_t(\psi)-2 Y_t ] }{2\tilde\nu^*_t(\gamma)} + \frac{[ \nu^*_t(\gamma) -  \tilde\nu^*_t(\gamma) ] [Y_t - \lambda_t(\psi) ]^2}{2 \nu^*_t(\gamma) \tilde\nu^*_t(\gamma)}  }  \\
	& \quad \quad +
	\frac{1}{2} \sup_{\gamma \in \Gamma} \norm{\log \frac{\tilde\nu^*_t(\gamma)}{\nu^*_t(\gamma)}}  \\
	& \leq \frac{1}{\underline{\nu}^*} a_t \Big( a_t + \norm{Y_t} +  \sup_{\psi \in \Psi} \norm{\lambda_t(\psi)} \Big) + \frac{1}{\underline{\nu}^{*2}} b_t \Big( Y^2_t + \sup_{\psi \in \Psi}\lambda^2_t(\psi) \Big) \\
	& \quad \quad + \frac{1}{2} \sup_{\gamma \in \Gamma} \norm{\log \left(  1 +  \frac{\tilde\nu^*_t(\gamma) - \nu^*_t(\gamma) }{\nu^*_t(\gamma)} \right) } \\
	& \leq \frac{1}{\underline{\nu}^*} a_t \Big( 1 + \norm{Y_t} +  \sup_{\psi \in \Psi} \norm{\lambda_t(\psi)} \Big) + \frac{1}{\underline{\nu}^{*2}} b_t \Big( Y^2_t + \sup_{\psi \in \Psi}\lambda^2_t(\psi) \Big) +  \frac{1}{2\underline{\nu}^*} b_t,
\end{align*}
for $t$ large enough since, by assumption \ref{Ass. starting value}, a.s. $a_t \to 0$ as $t \to\infty$. Note that in the last inequality we have used the fact that $x/(x+1) \leq \log(1+x) \leq x$ for $ x > -1 $ and that $|\log(1+x)| \leq \max\left\lbrace |x/(x+1)|, |x| \right\rbrace$. Indeed, by setting the simplified notation $x = (\tilde \nu- \nu)/\nu$, it is clear that $x= \tilde \nu/\nu -1>-1 $ since $\tilde \nu/\nu>0$. 
	By standard algebra we find that $|x/(x+1)|=|\tilde \nu - \nu|/\tilde \nu$. Therefore $|\log(1+x)| \leq \max\left\lbrace |\tilde \nu - \nu|/\tilde \nu, |\tilde \nu - \nu|/\nu \right\rbrace \leq b_t/ \underline{\nu}^*$ where the last inequality follows by \ref{Ass. lower bound} and the definition of $b_t$. Assumption \ref{Ass. starting value}  and an application of Cesaro's lemma 
lead to 
\begin{equation*}
	\sup_{\theta \in \Theta} |\tilde{L}_T(\theta) - L_T(\theta)| \leq	T^{-1}\sum_{t=1}^{T} \sup_{\theta \in \Theta} |\tilde{l}_t(\theta) - l_t(\theta)| \to 0\,, \quad a.s.
\end{equation*}
as $T\to\infty$. \hfill$\square$\\

\begin{lemma}
	\label{lemma2}
	Consider the PVQMLE in  \eqref{unconstrained} with score \eqref{score}. Under conditions~\ref{Ass. lower bound}-\ref{Ass. starting value} and \ref{Ass. starting value derivative}, almost surely as $T \to \infty$,
	$ \sqrt{T} \sup_{\theta \in \Theta} \|\tilde{S}_T(\theta)- S_T(\theta)\| \to 0 $.

\end{lemma}

\noindent \textbf{Proof of Lemma~\ref{lemma2}:} We obtain that
\begin{align*}
	\sup_{\theta \in \Theta} \lnorm{s_t(\theta)- \tilde{s}_t(\theta)}
	\leq & \sup_{\theta \in \Theta} \lnorm{\frac{1}{2\tilde\nu^*_t(\gamma)} \frac{\partial \tilde\nu^*_t(\gamma)}{\partial \theta} - \frac{1}{2\nu^*_t(\gamma)} \frac{\partial \nu^*_t(\gamma)}{\partial \theta}} \\
	&  + \sup_{\theta \in \Theta} \lnorm{\frac{Y_t-\tilde{\lambda}_t(\psi)}{\tilde\nu^*_t(\gamma)} \frac{\partial \tilde{\lambda}_t(\psi)}{\partial \theta} - \frac{Y_t-\lambda_t(\psi)}{\nu^*_t(\gamma)} \frac{\partial \lambda_t(\psi)}{\partial \theta}} \\
	&  + \sup_{\theta \in \Theta} \lnorm{\frac{ [ Y_t-\tilde{\lambda}_t(\psi)] ^2}{2\tilde\nu^{*2}_t(\gamma)} \frac{\partial \tilde\nu^*_t(\gamma)}{\partial \theta} - \frac{[ Y_t-\lambda_t(\psi)] ^2}{2\nu^{*2}_t(\gamma)} \frac{\partial \nu^*_t(\gamma)}{\partial \theta}} = \delta^1_t + \delta^2_t + \delta^3_t,
\end{align*}
with obvious notation. We now bound the single terms individually. In what follows the notation $o(1)$ almost surely, as $t \to \infty$, will be abbreviated to $o(1)$. 
\begin{align*}
	\delta^1_t &\leq \sup_{\theta \in \Theta} \lnorm{\frac{1}{2\tilde\nu^*_t(\gamma)}\left( \frac{\partial \tilde\nu^*_t(\gamma)}{\partial \theta} -  \frac{\partial \nu^*_t(\gamma)}{\partial \theta} \right) + \frac{\left[ \nu^*_t(\gamma) - \tilde\nu^*_t(\gamma) \right] }{2\tilde\nu^*_t(\gamma)\nu^*_t(\gamma)} \frac{\partial \nu^*_t(\gamma)}{\partial \theta} } \\ &\leq \frac{d_t}{2 \underline{\nu}^*} + \frac{b_t}{2 \underline{\nu}^{*2}} \sup_{\theta \in \Theta} \lnorm{\frac{\partial \nu^*_t(\gamma)}{\partial \theta}}\,.
\end{align*} 
Similarly,
\begin{align*}
	\delta^2_t &\leq \sup_{\theta \in \Theta} \lnorm{\frac{Y_t-\tilde{\lambda}_t(\psi)}{\tilde\nu^*_t(\gamma)} \left( \frac{\partial \tilde{\lambda}_t(\psi)}{\partial \theta} - \frac{\partial \lambda_t(\psi)}{\partial \theta} \right)} \\
	& \quad \quad + \sup_{\theta \in \Theta} \lnorm{  \frac{\partial \lambda_t(\psi)}{\partial \theta} \left( \frac{\lambda_t(\psi) - \tilde{\lambda}_t(\psi)}{\tilde\nu^*_t(\gamma)}+ \frac{Y_t-\lambda_t(\psi)}{\tilde\nu^*_t(\gamma)} - \frac{Y_t-\lambda_t(\psi)}{\nu^*_t(\gamma)} \right) } \\
	& \leq \frac{c_t}{ \underline{\nu}^*} \Big( \norm{Y_t} + \sup_{\psi \in \Psi} \norm{\lambda_t(\psi)} + a_t \Big) \\
	& \quad \quad + \sup_{\theta \in \Theta} \lnorm{\frac{\partial \lambda_t(\psi)}{\partial \theta} }  \left(   \frac{a_t}{\underline{\nu}^*} + \sup_{\theta \in \Theta} \norm{\frac{\left[ \nu^*_t(\gamma) - \tilde\nu^*_t(\gamma) \right] \left[ Y_t - \lambda_t(\psi) \right] }{\tilde\nu^*_t(\gamma)\nu^*_t(\gamma)  }    } \right) \\
	& \leq \frac{c_t}{ \underline{\nu}^*} \Big( \norm{Y_t} + \sup_{\psi \in \Psi} \norm{\lambda_t(\psi)} + o(1) \Big) + \sup_{\theta \in \Theta} \lnorm{\frac{\partial \lambda_t(\psi)}{\partial \theta} } \Big( \frac{a_t}{\underline{\nu}^*} + \frac{b_t}{\underline{\nu}^{*2}}\Big(  \norm{Y_t} + \sup_{\psi \in \Psi} \norm{\lambda_t(\psi)} \Big)   \Big) .
\end{align*}
Using similar arguments for $\delta^3_t$ and assumption \ref{Ass. starting value} leads to
\begin{align*}
	\delta^3_t &\leq  \frac{d_t}{\underline{\nu}^{*2}} \sup_{\theta \in \Theta}\left( Y_t^2 +  \lambda^2_t(\psi) + a_t^2 + 2 a_t \lambda_t(\psi)  \right) \\
	& \quad \quad +  \sup_{\theta \in \Theta} \lnorm{\frac{\partial \nu^*_t(\gamma)}{\partial \theta} } \sup_{\theta \in \Theta} \norm{\frac{[ \tilde{\lambda}_t(\psi) - \lambda_t(\psi)][  \tilde{\lambda}_t(\psi)+\lambda_t(\psi)-2 Y_t ] }{2\tilde\nu^{*2}_t(\gamma)}   }\\
	& \quad \quad + \sup_{\theta \in \Theta} \lnorm{\frac{\partial \nu^*_t(\gamma)}{\partial \theta} } \sup_{\theta \in \Theta} \norm{\frac{\left[ \nu^*_t(\gamma) -\tilde\nu^*_t(\gamma)\right] \left[ \nu^*_t(\gamma) + \tilde\nu^*_t(\gamma)\right] \left[ Y_t - \lambda_t(\psi) \right] ^2 }{2\nu^{*2}_t(\gamma)\tilde\nu^{*2}_t(\gamma)}}\\
	&\leq \frac{d_t}{\underline{\nu}^{*2}} \Big( Y_t^2 + \sup_{\psi \in \Psi} \lambda^2_t(\psi) + o(1)  \Big) + \sup_{\theta \in \Theta} \lnorm{\frac{\partial \nu^*_t(\gamma)}{\partial \theta} } \frac{a_t}{\underline{\nu}^{*2}} \left(  \norm{Y_t} + \sup_{\psi \in \Psi} \norm{\lambda_t(\psi)} + o(1) \right) \\
	& \quad \quad +   \sup_{\theta \in \Theta} \lnorm{\frac{\partial \nu^*_t(\gamma)}{\partial \theta} } \frac{2 b_t}{\underline{\nu}^{*3}} \Big( Y_t^2 + \sup_{\psi \in \Psi} \lambda^2_t(\psi) \Big) \,.
\end{align*}
By assumption \ref{Ass. starting value derivative}, $\delta^j_t = \Ob(t^{-\delta})$, for $\delta>1/2$ and $j=1,2,3$. Therefore  $\sqrt{T} \sup_{\theta \in \Theta} \|S_T(\theta)- \tilde{S}_T(\theta)\| \leq T^{-1/2} \sum_{t=1}^{T} \Ob(t^{-\delta}) $ converges to $0$ almost surely as $T \to \infty$.  \hfill$\square$\\

	\begin{lemma}  \label{Lem invertibility}
		Assumption~\ref{Ass. hessians} is satisfied for the unrestricted PVQMLE \eqref{unconstrained} under the following sufficient condition. 
		\begin{enumerate}[label=\textbf{A9$^*$}]
			\item \label{Ass. linear indipendence}
			The random variables of the vectors $\partial \lambda_t(\psi_0)/\partial \psi$ and  $\partial \nu^*_t(\gamma^*)/\partial \gamma$ are linearly independent. Moreover, one of the following conditions holds a.s. for some $t\geq 1$.
		\begin{enumerate}[label=\arabic*.]
				\item $\nu_t^*(\gamma^*)=\nu_t$.
				\item $\nu_t^*(\gamma^*) < \nu_t$, ~~ $\partial^2 \nu_t^*(\gamma^*)/\partial \theta \partial \theta'$ is negative semi-definite.
				\item $\nu_t < \nu_t^*(\gamma^*) < 2 \nu_t$, ~~ $\partial^2 \nu_t^*(\gamma^*)/\partial \theta \partial \theta'$ is positive semi-definite.
				\item $\nu_t^*(\gamma^*) < 2 \nu_t$, ~~~~ $\partial^2 \nu_t^*(\gamma^*)/\partial \theta_i \partial \theta_j=0$ for all $i,j=1,\dots,m$.
			\end{enumerate}
		\end{enumerate}
		The same result holds for the restricted PVQMLE \eqref{constrained} with $\partial \gamma_2$ instead of $\partial \gamma$.
	\end{lemma}

\noindent \textbf{Proof of Lemma~\ref{Lem invertibility}:}  Condition~\ref{Ass. hessians} requires that for all $ \eta \in \R^m$, $\eta' \, \E [-\partial^2 l_t(\theta_0)/\partial \theta \partial \theta']  \eta > 0$,  with $ \eta \neq 0$, but $\E [-\partial^2 l_t(\theta_0)/\partial \theta \partial \theta']=\E [\E[-\partial^2 l_t(\theta_0)/\partial \theta \partial \theta'|\Fb_{t-1}]]$ and following \eqref{second derivative}, we only need to show 
	\begin{equation} \label{hessian pos def}
		\E \left(  d_t \eta'  f_t^\theta f_t^{\theta\prime} \eta + l_t \eta'  h_t^\theta h_t^{\theta\prime} \eta + \eta' C^\theta_t \eta \right)    >0 \,,
	\end{equation}
	where
	\begin{equation*}
		d_t = \frac{1}{\nu^*_t(\gamma^*)}\,, ~~~  f_t^\theta =  \frac{\partial \lambda_t(\psi_0)}{\partial \theta}\,, ~~~ l_t =  \frac{2\nu_t-\nu^*_t(\gamma^*)}{2\nu^{*3}_t(\gamma^*)}  \,, ~~~ h_t^\theta = \frac{\partial \nu^*_t(\gamma^*)}{\partial \theta}\,, ~~~ C^\theta_t =  \left( \frac{\nu^*_t(\gamma^*) - \nu_t}{2\nu^{*2}_t(\gamma^*)}  \right)   \frac{\partial^2 \nu^*_t(\gamma^*)}{\partial \theta \partial \theta^\prime}\,.
	\end{equation*}
	Note that under the conditions in \ref{Ass. linear indipendence} we have that a.s. $l_t >0$ and $\eta' C^\theta_t \eta \geq 0$. Moreover, a.s.
	$d_t>0$,  
	$\eta' f_t^\theta f_t^{\theta\prime} \eta = (\eta' \, f_t^\theta)^2 \geq 0 $ and $\eta' h_t^\theta h_t^{\theta\prime} \eta = (\eta' \, h_t^\theta)^2 \geq 0 $. Therefore a sufficient condition for \eqref{hessian pos def} requires a.s. $\eta' \, f_t^\theta \neq 0$ or $\eta' h_t^\theta \neq 0 $.
	Let $0_{m}$ be a $m$-dimensional vector of zeros. To prove the result recall that 
	$$
	f_t^\theta =
	\begin{pmatrix}
		f_t^\psi  \\
		0_{k}
	\end{pmatrix}\,, ~~~~
	h^\theta_t = \begin{pmatrix}
		h_t^\psi\\
	h_t^\gamma
	\end{pmatrix}\,.
	$$
	Hence, a.s.
	$\eta' \, f_t^\theta = \eta_1' \, f_t^\psi $.
	We can split $\eta = (\eta_1', \eta_2')'$ where $\eta_1$ has dimension $p$ and $\eta_2$ has dimension $k$. Consider two cases: (i) $\eta_1 \neq 0$ and (ii) $\eta_1 = 0$. Under (i) the result is verified by a.s. $\eta_1' \, f_t^\psi \neq 0$. In the case (ii), the result follows by $ \eta_2' \, h_t^\gamma \neq 0$ a.s.  since $\eta' \, h_t^\theta = \eta_2' \, h_t^\gamma $.

	Recall that $\eta'I(\theta_0)\eta = \E[(\eta's_t(\theta_0))^2] \geq 0 $. Therefore, to prove the positive definiteness of $I(\theta_0)$  we need to show that for all $ \eta \in \R^m$, with $ \eta \neq 0$, $\eta's_t(\theta_0) \neq 0$ where
	\begin{equation} \label{fisher pos def}
		\eta's_t(\theta_0) = \frac{e_t}{\nu^*_t(\gamma^*)} \eta' f_t^\theta + \frac{e_t^2 - \nu^*_t(\gamma^*)}{2 \nu^{*2}_t(\gamma^*)}  \eta'  h_t^\theta\,,
	\end{equation}
	and $e_t = Y_t - \lambda_t(\psi_0)$ and therefore a.s. $e_t \neq 0$, $e_t^2 - \nu^*_t(\gamma^*) \neq 0$ and $\nu^*_t(\gamma^*) > 0$. If only one between $ \eta' f_t^\theta$ and $ \eta' h_t^\theta$ is different from $0$ a.s. then the result follows with argument identical to the Hessian matrix case above. In the case where both $ \eta' f_t^\theta$ and $ \eta' h_t^\theta$ are not $0$ a.s. we show that this cannot imply that $\eta's_t(\theta_0) = 0$ since by equation~\eqref{fisher pos def} this would entail 
	\begin{equation*}
		\eta' f_t^\theta = -\frac{e_t^2 - \nu^*_t(\gamma^*)}{2 \nu^*_t(\gamma^*)e_t}  \eta'  h_t^\theta\,,
	\end{equation*}
	where the left-hand side is $\Fb_{t-1}$-measurable whereas the right-hand side is not as it depends on $Y_t$. Therefore  $\eta's_t(\theta_0) \neq 0$ for any non-trivial vector $\eta$. This concludes the proof.
	 \hfill$\square$

\section{Further numerical results}

\label{Appendix B further numerical results}

\subsection{Outliers}
We evaluate the robustness of the described test statistic by repeating the same simulation study as in Section \ref{SEC: Specification tests} with the inclusion of an outlier  defined as 3 times the standard deviation of the observations plus their sample mean. The results are summarized in Tables~\ref{Tab: empirical size pois vs nb outlier}-\ref{Tab: estimation outlier} and Figure~\ref{Fig: power pois vs nb outlier} below.   

	\begin{table}[h!]
	\centering
	\caption{Empirical size for test in  \eqref{test pois thin} with outlier.
		The model considered under $H_0$ is an INAR($1$) model with Poisson thinning as well as Poisson error with parameter values $a=0.75$ and  $\omega=1$.}
	\scalebox{0.95}{
		\begin{tabular}{cccccc}
			\hline
			&	\multicolumn{4}{c}{$T$} \\
			\cline{2-6}
			Nominal size & 100 & 250 & 500 & 1000 & 2000 \\ 
			\noalign{\smallskip}\hline \noalign{\smallskip}
			0.1000	& 0.1510 & 0.1544 & 0.1312 & 0.1154 & 0.1112 \\ 
			0.0500 & 0.0692	& 0.0730 & 0.0646 & 0.0550 & 0.0540 \\ 
			0.0100 & 0.0142	& 0.0116 & 0.0106 & 0.0112 & 0.0128 \\ 
			\hline
		\end{tabular}
	}
	\label{Tab: empirical size pois vs nb outlier}
\end{table}

	\begin{table}[h!]
	\centering
	\caption{Mean of parameters estimated for unrestricted PVQMLE over 5000 simulations with the presence of outlier.
		Data are generated from an INAR($1$) model with Poisson thinning as well as Poisson error with parameter values $a=0.75$ and  $\omega=1$.}
	\scalebox{0.95}{
		\begin{tabular}{cccccc}
		  \hline
		$T$ & $a$ & $\omega_1$ & $\omega_2$ & $b$ \\ 
			\noalign{\smallskip}\hline \noalign{\smallskip}
		
		100 & 0.6827 & 1.2679 & 1.7522 & 0.8230 \\ 
		250 & 0.7235 & 1.1059 & 1.3657 & 0.7740 \\ 
		500 & 0.7372 & 1.0497 & 1.1805 & 0.7616 \\ 
		1000 & 0.7432 & 1.0268 & 1.1066 & 0.7509 \\ 
		2000 & 0.7467 & 1.0125 & 1.0562 & 0.7497 \\ 
		
			\hline
		\end{tabular}
	}
	\label{Tab: estimation outlier}
\end{table}

\begin{figure}[h!]
	\centering
	\begin{tabular}{c}
		\includegraphics[width=0.8\linewidth]{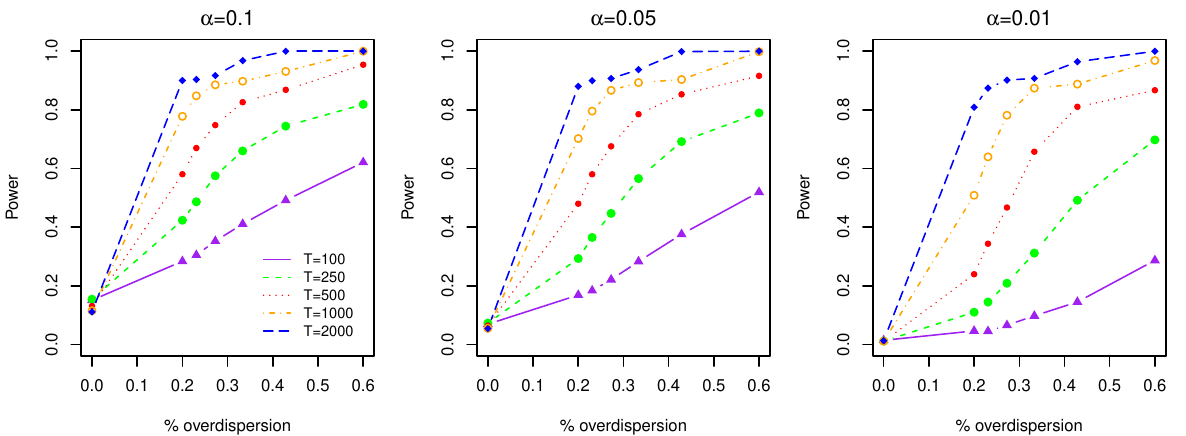}
	\end{tabular}	
	\caption{Empirical power for test in \eqref{test pois thin} with outlier. The true parameter values of the  INAR($1$) model with negative binomial thinning and Poisson error are $a=0.75$ and $\omega = 1$. The value of the dispersion parameter $v$ changes  as indicated in the horizontal axis through the \% of overdispersion: $1-a/(a+a^2/v)$.}
	\label{Fig: power pois vs nb outlier}
\end{figure}

\subsection{Near unit root}
 We evaluate the test in case of near unit root by considering the same simulation setting as   in Section \ref{SEC: Specification tests} but setting the autoregressive parameter equal to 0.99. The results are summarized in Tables~\ref{Tab: empirical size pois vs nb non-stationary}-\ref{Tab: estimation non-stationary} and  Figure~\ref{Fig: power pois vs nb non-stationary} below.

	\begin{table}[h!]
	\centering
	\caption{Empirical size for test in  \eqref{test pois thin}.
		The model considered under $H_0$ is an INAR($1$) model with Poisson thinning as well as Poisson error with parameter values $a=0.99$ and  $\omega=1$.}
	\scalebox{0.95}{
		\begin{tabular}{cccccc}
			\hline
			&	\multicolumn{4}{c}{$T$} \\
			\cline{2-6}
			Nominal size & 100 & 250 & 500 & 1000 & 2000 \\ 
			\noalign{\smallskip}\hline \noalign{\smallskip}
		0.1000 & 0.0524 & 0.0562 & 0.0568 & 0.0614 & 0.0660 \\ 
		0.0500 & 0.0250 & 0.0310 & 0.0300 & 0.0296 & 0.0254 \\ 
		0.0100& 0.0074 & 0.0060 & 0.0076 & 0.0068 & 0.0050 \\ 
			\hline
		\end{tabular}
	}
	\label{Tab: empirical size pois vs nb non-stationary}
\end{table}

	\begin{table}[h!]
	\centering
	\caption{Mean of parameters estimated for unrestricted PVQMLE over 5000 simulations with the presence of outlier.
		Data are generated from an INAR($1$) model with Poisson thinning as well as Poisson error with parameter values $a=0.99$ and  $\omega=1$.}
	\scalebox{0.95}{
		\begin{tabular}{cccccc}
			\hline
			$T$ & $a$ & $\omega_1$ & $\omega_2$ & $b$ \\ 
			\noalign{\smallskip}\hline \noalign{\smallskip}
			
		100 & 0.9364 & 6.0910 & 16.8541 & 0.7962 \\ 
		250 & 0.9690 & 2.9188 & 6.8376 & 0.9150 \\ 
		500 & 0.9802 & 1.8379 & 3.3397 & 0.9578 \\ 
		1000 & 0.9857 & 1.3415 & 1.7776 & 0.9776 \\ 
		2000 & 0.9880 & 1.1619 & 1.1795 & 0.9866 \\ 
			
			\hline
		\end{tabular}
	}
	\label{Tab: estimation non-stationary}
\end{table}

\begin{figure}[h!]
	\centering
	\begin{tabular}{c}
		\includegraphics[width=0.8\linewidth]{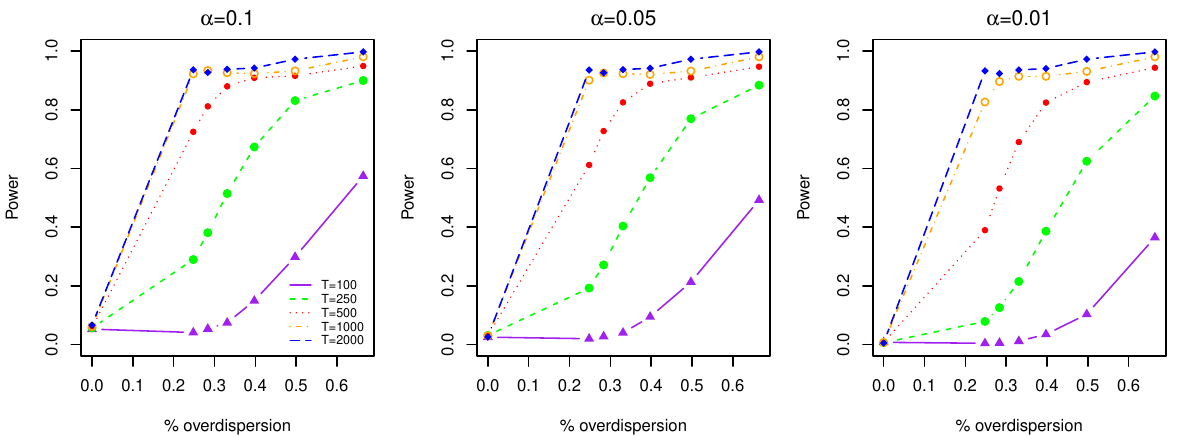}
	\end{tabular}	
	\caption{Empirical power for test in \eqref{test pois thin}. The true parameter values of the  INAR($1$) model with negative binomial thinning and Poisson error are $a=0.99$ and $\omega = 1$. The value of the dispersion parameter $v$ changes  as indicated in the horizontal axis through the \% of overdispersion: $1-a/(a+a^2/v)$.}
	\label{Fig: power pois vs nb non-stationary}
\end{figure}

\subsection{BiNB thinning}
We  evaluate the power of the equidispersion test under BiNB  thinning.  In this specification $X_j \sim BerG(\mu, \pi)$ is the Bernoulli-Geometric distribution, defined as the sum of a Bernoulli distribution with probability $\pi$ and a geometric distribution with mean $\mu$ where the distributions are independent and $\mu+\pi <1$. Moreover, $a=\mu+\pi$, $b=\pi(1-\pi)+\mu(1-\mu)$ so $b/a=1+\mu-\pi$ and therefore $b>a$ if $\mu>\pi$. Figure~\ref{Fig: power pois vs binb} below shows the power of the test in \eqref{test pois thin} to reject the null hypothesis. 

\begin{figure}[h!]
	\centering
	\begin{tabular}{c}
		\includegraphics[width=0.8\linewidth]{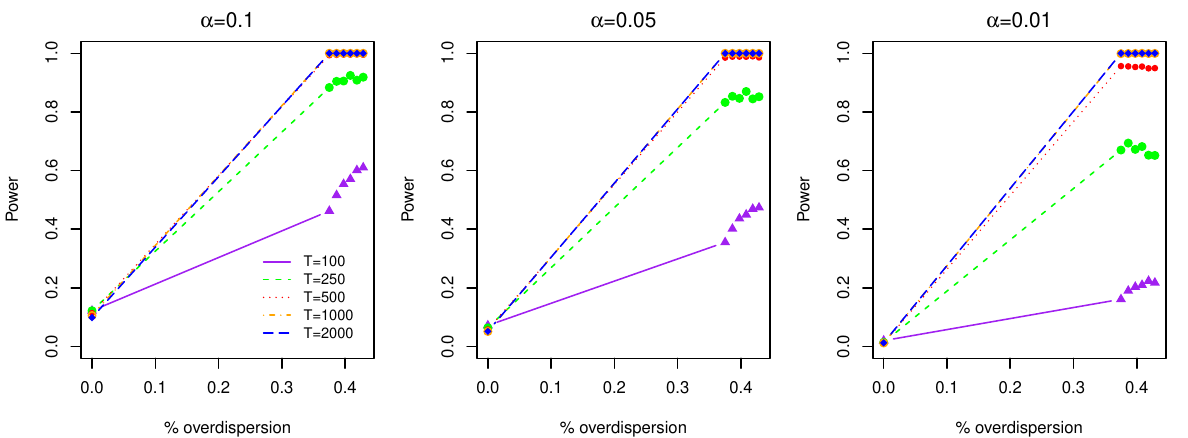}
	\end{tabular}	
	\caption{Empirical power for test in \eqref{test pois thin}. The true parameter values of the  INAR($1$) model with BiNB thinning and Poisson error are $\mu=0.75$ and $\omega = 1$. The value of the Bernoulli parameter $\pi\in[0, 0.15]$ changes  as indicated in the horizontal axis through the \% of overdispersion: $1-1/(1+\mu-\pi)$.}
	\label{Fig: power pois vs binb}
\end{figure}

\subsection{INAR($2$)}
We consider the case of testing equidispersion of the thinning operator in an INAR(2) model. In this case the hypothesis test is the following
\begin{equation} \label{test pois thin 2}
	H_0: b_i = a_i  \quad  \text{vs} \quad H_1: b_i \neq a_i \quad \text{for} ~ i=1,2\,.
\end{equation}
The results of the test against negative binomial thinning are reported in Tables~\ref{Tab: empirical size pois vs nb 2}-\ref{Tab: estimation 2} and Figure~\ref{Fig: power pois vs nb 2} below. 

\begin{table}[h!]
	\centering
	\caption{Empirical size for test in  \eqref{test pois thin 2}.
		The model considered under $H_0$ is an INAR($2$) model with Poisson thinning as well as Poisson error with parameter values $a_1=a_2=0.4$ and  $\omega=1$.}
	\scalebox{0.95}{
		\begin{tabular}{cccccc}
			\hline
			&	\multicolumn{4}{c}{$T$} \\
			\cline{2-6}
			Nominal size & 100 & 250 & 500 & 1000 & 2000 \\ 
			\noalign{\smallskip}\hline \noalign{\smallskip}
			0.1000 & 0.1384 & 0.1460 & 0.1348 & 0.1182 & 0.1096 \\ 
			0.0500 & 0.0932 & 0.0822 & 0.0800 & 0.0632 & 0.0574 \\ 
			0.0100 & 0.0452 & 0.0242 & 0.0206 & 0.0166 & 0.0138 \\ 
			\hline
		\end{tabular}
	}
	\label{Tab: empirical size pois vs nb 2}
\end{table}

\begin{table}[h!]
	\centering
	\caption{Mean of parameters estimated for unrestricted PVQMLE over 5000 simulations.
		Data are generated from an INAR($2$) model with Poisson thinning as well as Poisson error with parameter values $a_1=a_2=0.4$ and  $\omega=1$.}
	\scalebox{0.95}{
		\begin{tabular}{ccccccc}
			\hline
			$T$ & $a_1$ & $a_2$ & $\omega_1$ & $\omega_2$ & $b_1$ & $b_2$ \\ 
			\noalign{\smallskip}\hline \noalign{\smallskip}
			
			100 & 0.3835 & 0.3614 & 1.2498 & 1.0030 & 0.3939 & 0.3865 \\ 
			250 & 0.3937 & 0.3863 & 1.0858 & 0.9856 & 0.3978 & 0.3964 \\ 
			500 & 0.3966 & 0.3942 & 1.0386 & 0.9838 & 0.3997 & 0.3986 \\ 
			1000 & 0.3976 & 0.3977 & 1.0189 & 0.9977 & 0.3979 & 0.3991 \\ 
			2000 & 0.3991 & 0.3984 & 1.0099 & 0.9989 & 0.3999 & 0.3991 \\ 
			
			\hline
		\end{tabular}
	}
	\label{Tab: estimation 2}
\end{table}

\begin{figure}[h!]
	\centering
	\begin{tabular}{c}
		\includegraphics[width=0.8\linewidth]{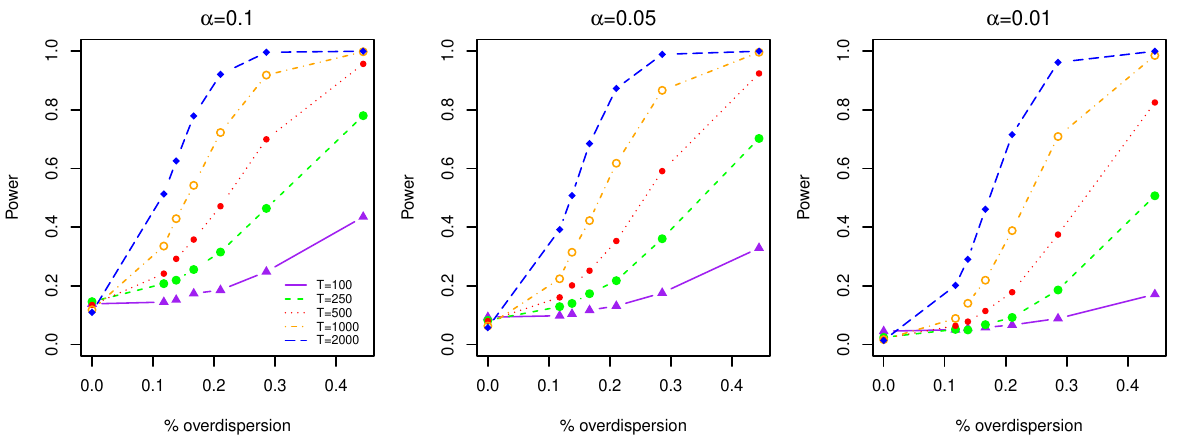}
	\end{tabular}	
	\caption{Empirical power for test in \eqref{test pois thin 2}. The true parameter values of the  INAR($2$) model with negative binomial thinning and Poisson error are $a_1=a_2=0.4$ and $\omega = 1$. The value of the dispersion parameter $v$ changes  as indicated in the horizontal axis through the \% of overdispersion: $1-a_1/(a_1+a_1^2/v)$. The horizontal axis is equal to $1-a_2/(a_2+a_2^2/v)$ since $a_1=a_2$.}
	\label{Fig: power pois vs nb 2}
\end{figure}

\end{appendices}

\bibliographystyle{apalike}
\bibliography{references}

\end{document}